\documentclass[12pt]{article}
\usepackage{amsmath,amssymb}
\usepackage{bm}
\usepackage{color}
\usepackage{cite}
\usepackage{mathtools}

\def\slash#1{\not\!\!#1}

\begin{document}

\title{
\begin{flushright}
\ \\*[-80pt]
\begin{minipage}{0.2\linewidth}
\normalsize
EPHOU-20-008\\
KEK-TH-2239\\*[50pt]
\end{minipage}
\end{flushright}
{\Large \bf
Modular symmetry by orbifolding magnetized $T^2\times T^2$:\\
realization of double cover of $\Gamma_N$ 
\\*[20pt]}}

\author{
Shota Kikuchi$^a$,
~Tatsuo Kobayashi$^a$,
~Hajime Otsuka$^b$, \\
~Shintaro Takada$^a$,
 and~Hikaru Uchida$^a$
\\*[20pt]
\centerline{
\begin{minipage}{\linewidth}
\begin{center}
{\it \normalsize
$^a$Department of Physics, Hokkaido University, Sapporo 060-0810, Japan \\
$^b$KEK Theory Center, Institute of Particle and Nuclear Studies, KEK \\
1-1 Oho, Tsukuba, Ibaraki 305-0801, Japan} \\*[5pt]
\end{center}
\end{minipage}}
\\*[50pt]}

\date{
\centerline{\small \bf Abstract}
\begin{minipage}{0.9\linewidth}
\medskip
\medskip
\small
We study the modular symmetry of zero-modes on $T_1^2 \times T_2^2$ and orbifold compactifications with 
magnetic fluxes, $M_1,M_2$, 
where modulus parameters are identified.
This identification breaks  the modular symmetry of $T^2_1 \times T^2_2$, $SL(2,\mathbb{Z})_1 \times SL(2,\mathbb{Z})_2$ 
 to $SL(2,\mathbb{Z})\equiv\Gamma$.
Each of the wavefunctions on $T^2_1 \times T^2_2$ and orbifolds behaves as the modular forms of weight 1 for the principal congruence subgroup $\Gamma$($N$), $N$ being 
2 times the least common multiple of $M_1$ and $M_2$.
Then, zero-modes transform each other under the modular symmetry as 
multiplets of double covering groups of $\Gamma_N$ such as the double cover of $S_4$.
\end{minipage}
}

\begin{titlepage}
\maketitle
\thispagestyle{empty}
\end{titlepage}

\newpage


\section{Introduction}
\label{Intro}

The origin of the flavor structure such as the masses and the mixing angles of quarks and leptons is one of the significant mysteries of the standard model (SM).
Many ideas have been proposed to understand the flavor structure.
Among them, non-Abelian discrete flavor models~\cite{
	Altarelli:2010gt,Ishimori:2010au,Ishimori:2012zz,Hernandez:2012ra,
	King:2013eh,King:2014nza,Tanimoto:2015nfa,King:2017guk,Petcov:2017ggy} are attractive.
In such flavor models, various non-Abelian discrete symmetries such as $S_N$, $A_N$, $\Delta(3N^2)$, $\Delta(6N^2)$ are assumed as  symmetries of  quark and lepton flavors.
In those models, the realistic masses and mixing angles of quarks or leptons are obtained through breaking the flavor symmetries by vacuum expectation values (VEVs) of gauge singlet scalars, so-called flavons.
However, a complicate vacuum alignment is required.

Extra dimensional theory such as superstring theory 
can lead to non-Abelian discrete flavor symmetries as geometrical symmetries. (See Refs.~\cite{Kobayashi:2006wq,Abe:2009vi}.)
In particular, the two-dimensional (2D) torus $T^2$ and orbifolds have the geometrical symmetry, the so-called modular symmetry, $\Gamma \equiv SL(2,Z)$ (or $\overline{\Gamma} \equiv SL(2,Z)/\mathbb{Z}_2$).
Zero-modes on such a geometry, corresponding to flavors of the SM quarks or leptons, transform under the modular transformation. It was investigated in magnetized D-brane models \cite{Kobayashi:2017dyu,Kobayashi:2018rad,Kobayashi:2018bff,Ohki:2020bpo,Kikuchi:2020frp}
and heterotic orbifold models \cite{Lauer:1989ax,Lerche:1989cs,Ferrara:1989qb,Baur:2019kwi,Nilles:2020nnc}.
(See also \cite{Kobayashi:2016ovu,Kariyazono:2019ehj,Kobayashi:2020hoc}.)
In this sense, the modular symmetry is regarded as a flavor symmetry.
In particular, Ref~\cite{Kikuchi:2020frp} shows that $|M|$ zero-mode wavefunctions on $T^2$ with magnetic flux $M$ behave as modular forms of weight $1/2$ for $\widetilde{\Gamma}(2|M|)$, which is a normal subgroup of the double covering group of $\Gamma$, i.e. $\widetilde{\Gamma} \equiv \widetilde{SL}(2,Z)$, and then they are representations of the quotient group $\widetilde{\Gamma}'_{2|M|} \equiv \widetilde{\Gamma}/\widetilde{\Gamma}(2|M|)$.
It is notable that the Yukawa couplings as well as higher order couplings also transform non-trivially under the modular transformation.
In addition, instead of VEVs of flavons, the flavor symmetry coming from the modular symmetry is broken when the modulus is fixed through the modulus stabilization.

It is also interesting that the finite modular groups $\Gamma_N \equiv \overline{\Gamma}/\overline{\Gamma}(N)$ for $N=2,3,4,5$ are isomorphic to $S_3$, $A_4$, $S_4$, $A_5$~\cite{deAdelhartToorop:2011re}, respectively.
Note that $\overline{\Gamma}(N)$ ($\Gamma(N)$) is a normal subgroup of $\overline{\Gamma}$ ($\Gamma$), so-called the principal congruence subgroup of level $N$.
Recently, a lot of bottom-up approaches of flavor models inspired by these aspects have been studied~\cite{Feruglio:2017spp,Kobayashi:2018vbk,Penedo:2018nmg,Kobayashi:2018scp,deAnda:2018ecu,Okada:2018yrn,Ding:2019xna,
Nomura:2019jxj,Novichkov:2019sqv,
Asaka:2019vev,Zhang:2019ngf,Wang:2019ovr,Kobayashi:2019gtp,Lu:2019vgm,Wang:2019xbo,Abbas:2020qzc}.
In those models, the Yukawa couplings as well as higher order couplings are treated as modular forms of even-number weights.
Furthermore, Ref.~\cite{Liu:2019khw} shows modular forms of odd-number weights are representations of $\Gamma'_N \equiv \Gamma/\Gamma(N)$, which is the double covering group of $\Gamma_N$.
In the latest studies~\cite{Novichkov:2020eep,Liu:2020akv}, flavor models of $\Gamma'_4\simeq S_4'$ with modular forms of weight integer 
were studied.
Thus, it is important to study the modular flavor symmetry, $\Gamma_N$ and its covering groups  $\Gamma'_N$from 
both top-down and bottom-up approaches.

The moduli stabilization is a key issue. 
Three-form fluxes can  stabilize  complex structure modulus~\cite{Giddings:2001yu} as well as the dilaton.
In Ref.~\cite{Kobayashi:2020hoc}, for example, $T^2_1 \times T^2_2 \times T^2_3$ with three-form fluxes have been considered and the complex modulus of $T^2_1$, $\tau_1$ and that of $T^2_2$, $\tau_2$ have to be related each other, such as $\tau_1 = \tau_2 \equiv \tau$ due to the three-form fluxes.
In other words, the modular symmetry on $T^2_1 \times T^2_2$, $\Gamma \times \Gamma$ is broken to $\Gamma$ due to the three-form fluxes below the heavy mass scale of the stabilized moduli.
A similar breaking $\Gamma \times \Gamma \to \Gamma$ can be realized by imposing a permutation symmetry 
between  $T^2_1$ and $T^2_2$.
Such a setup is quite interesting as follows.
The zero-mode wavefunctions on $T^2$ with magnetic flux $M$ 
behaves as the modular forms of the weight $1/2$ representing $\widetilde{\Gamma}'_{2|M|}$.
Thus we expect that the above setup would lead to zero-mode wavefunctions behaving 
the modular forms of the weight 1 and representing double covering  groups of $\Gamma_N$, i.e., $\Gamma'_N$.

Our purpose in this paper is to study the modular symmetry of zero-modes on 
 $T^2_1 \times T^2_2$ with magnetic fluxes, where the complex structure moduli are 
identified as $\tau_1=\tau_2=\tau$.
Furthermore, we also study its orbifolding by the $\mathbb{Z}_2$ twist, the $\mathbb{Z}_N$ shift, and also the $\mathbb{Z}_2$ 
permutation, which interchanges $T^2_1$ and $T^2_2$.
These orbifoldings decompose a representation to smaller representations such as irreducible representations.

This paper is organized as follows.
In section \ref{waveT2} 
we briefly review the zero-mode wavefunctions on $T^2$ with magnetic flux.
In section \ref{SymT2}, we give a review on the modular symmetry of zero-modes on $T^2$ and its orbifolding.
In section \ref{SymT2xT2}, we apply them on a magnetized $T^2_1 \times T^2_2$ and its orbifolding by the $\mathbb{Z}_2$ twist, the $\mathbb{Z}_N$ shift, and the $\mathbb{Z}_2$ permutation.
Here we identify $\tau_1 = \tau_2 \equiv \tau$.
We find that the wavefunctions on the $T^2_1 \times T^2_2$ behave as modular forms of weight 1 for 
$\Gamma(N)$.
The zore-modes are multiplets of $\Gamma_N'$ and $\Gamma_N$.
In section \ref{conclusion}, we conclude this study.


\section{Zero-mode wavefunctions on magnetized $T^2$}
\label{waveT2}

First, we review ten-dimensional (10D) ${\cal N}=1$ super Yang-Mills theory on $M^4 \times T^2_1 \times T^2_2 \times T^2_3$ with magnetic fluxes, which is the low energy effective field theory of superstring theory. 
Also, our setup in section \ref{SymT2xT2} is applicable to D7-brane models on $M^4 \times T^2_1 \times T^2_2$.
The 10D Lagrangian is given by
\begin{align}
S=\int_M d^4x \prod_{i=1,2,3} \int_{T^2_i} d^2z_i \left[-\frac{1}{4g^2} {\rm Tr}(F^{MN} F_{MN})+\frac{i}{2g} {\rm Tr}(\bar{\lambda}\Gamma^MD_M\lambda)\right],
\end{align}
where $F_{MN}=\partial_M A_N-\partial_N A_M-i[A_M,A_N]$ and $D_M\lambda=\partial_M\lambda-i[A_M,\lambda]$ with $M,N=0,1,...,9$.
By Kaluza-Klein decomposition, ten-dimensional vector potential $A_M$ and Majorana-Weyl spinors $\lambda$ can be written as 
\begin{align}
A_M(x,z_1,z_2,z_3) &= \sum_{n_1,n_2,n_3} \phi_{M,n_1n_2n_3}(x)\otimes\phi_{M,n_1}(z_1)\otimes\phi_{M,n_2}(z_2)\otimes\phi_{M,n_3}(z_3),\\
\lambda(x,z_1,z_2,z_3) &= \sum_{n_1,n_2,n_3} \psi_{n_1n_2n_3}(x)\otimes\psi_{n_1}(z_1)\otimes\psi_{n_2}(z_2)\otimes\psi_{n_3}(z_3).
\end{align}
Here $\psi_{n_i}(z_i)$ is the $n_i$-th excited mode of 2D Weyl spinors on the $i$-th torus, $T^2_i$, and satisfies the following Dirac equation,
\begin{align}
i\slash{D}_2 \psi_{n_i}(z_i)=m_{n_i}\psi_n(z_i). \label{2D massive Dirac}
\end{align}
Since $m_{n_i}$ gives the compact scale mass of the four-dimensional Weyl spinor,
we consider massless mode $\psi_0(z_i)$. In this section, we focus on one torus $T^2$, hence, we consider zero-mode wavefunctions on the magnetized $T^2$.

For simplicity, we study the wavefunctions on the torus $T^2$ with U(1) magnetic flux~\cite{Cremades:2004wa}.
The torus can be regarded as the complex plane $\mathbb{C}$ divided by a two-dimensional lattice $\Lambda$, that is $T^2\simeq \mathbb{C}/\Lambda$. 
Then, the lattice $\Lambda$ is characterized by the complex modulus parameter $\tau\equiv e_2/e_1 \ ({\rm Im}\tau>0)$, where $e_1$, $e_2$ are the basis that spans the lattice $\Lambda$. 
This torus has the metric such as
\begin{align}
ds^2 = 2 h_{\mu\nu} dz^{\mu} d\bar{z}^{\nu}, \quad
h = |e_1|^2
\begin{pmatrix}
0 & \frac{1}{2} \\
\frac{1}{2} & 0
\end{pmatrix},
\label{metric}
\end{align}
and the U(1) magnetic flux is given by
\begin{align}
F = \frac{i\pi M}{{\rm Im}\tau} dz \wedge d\bar{z}.  \label{F}
\end{align}
This flux leads to the vector potential one-form,
\begin{align}
A(z,\bar{z})
&= \frac{\pi M}{{\rm Im}\tau} {\rm Im}\left( (\bar{z}+\bar{\zeta})dz \right) \notag \\
&= - \frac{i\pi M}{2{\rm Im}\tau} (\bar{z}+\bar{\zeta}) dz + \frac{i\pi M}{2{\rm Im}\tau} (z+\zeta) d\bar{z} \label{A} \\
&= A_{z} dz + A_{\bar{z}} d\bar{z}, \notag
\end{align}
where $\zeta$ is a Wilson line phase. Since the complex coordinate $z$ on $T^2$ is identified with $z+1$ and $z+\tau$, the vector potential $A(z,\bar{z})$ obeys the following boundary conditions, 
\begin{align}
A(z+1,\bar{z}+1) &= A(z,\bar{z}) + d\left( \frac{\pi M}{{\rm Im}\tau} {\rm Im}z \right) = A(z,\bar{z}) + d\chi_1(z,\bar{z}), \label{z1A} \\
A(z+\tau,\bar{z}+\bar{\tau}) &= A(z,\bar{z}) + d\left( \frac{\pi M}{{\rm Im}\tau} {\rm Im}\bar{\tau}z \right) = A(z,\bar{z}) + d\chi_2(z,\bar{z}), \label{ztauA}
\end{align}
which correspond to $U(1)$ gauge transformation. Here, $\chi_1(z,\bar{z})$ and $\chi_2(z,\bar{z})$ are given by
\begin{align}
\chi_1(z,\bar{z}) = \frac{\pi M}{{\rm Im}\tau} {\rm Im}(z+\zeta),\ \chi_2(z,\bar{z}) = \frac{\pi M}{{\rm Im}\tau} {\rm Im}[\bar{\tau}(z+\zeta)].
\end{align}

Here and in what follows, we consider the 2D spinor with the U(1) charge $q=1$.
To preserve 2D Dirac equation, Eq.~(\ref{2D massive Dirac}), under U(1) gauge transformation, the boundary conditions for 2D Weyl spinors,
\begin{align}
\psi(z) =
\begin{pmatrix}
\psi^{M}_+(z) \\ \psi^{M}_-(z)
\end{pmatrix}, \quad
\psi^{M}_-(z) = \overline{\psi^{-M}_+(z)}, \label{spinor}
\end{align}
are obtained by
\begin{align}
&\psi(z+1) = e^{i\chi_1(z)} \psi(z) = e^{i\pi M \frac{{\rm Im}(z+\zeta)}{{\rm Im}\tau}} \psi(z), \label{psiz1} \\
&\psi(z+\tau) = e^{i\chi_2(z)} \psi(z) = e^{i\pi M \frac{{\rm Im}\bar{\tau}(z+\zeta)}{{\rm Im}\tau}} \psi(z). \label{psiztau}
\end{align}
Then, considering contractible loops on $T^2$, we obtain the Dirac quantization condition,
\begin{align}
M \in \mathbb{Z}.
\end{align}
To determine the zero-mode of $\psi(z)$, we need to compute the 2D Dirac operator,
\begin{align}
i\slash{D} &= i\gamma^z D_z +i\gamma^{\bar{z}} D_{\bar{z}} \\
 &= 
\begin{pmatrix}
0 & \frac{2i}{e_1} (\partial-iA_z) \\
\frac{2i}{\bar{e}_1} (\bar{\partial}-iA_{\bar{z}}) & 0
\end{pmatrix}
 = 
\begin{pmatrix}
0 & \frac{2i}{e_1} (\partial-\frac{\pi M}{2{\rm Im}\tau} (\bar{z}+\bar{\zeta})) \\
\frac{2i}{\bar{e}_1} (\bar{\partial}+\frac{\pi M}{2{\rm Im}\tau} (z+\zeta)) & 0
\end{pmatrix}, 
\end{align}
where $\gamma^{z}, \gamma^{\bar{z}}$ are given by
\begin{align}
\gamma^{z} = \frac{1}{e_1}
\begin{pmatrix}
0 & 2 \\
0 & 0
\end{pmatrix},
\quad
\gamma^{\bar{z}} = \frac{1}{\bar{e}_1}
\begin{pmatrix}
0 & 0 \\
2 & 0
\end{pmatrix},
\label{gammaz}
\end{align}
with $\{ \gamma^{z}, \gamma^{\bar{z}} \} = 2h^{z\bar{z}}$.
By solving the massless Dirac equation, $i\slash{D} \psi(z) = 0$,  for $M>0\ (M<0)$, we find only $\psi_+^M(z)\ (\psi_-^M(z))$ has the $|M|$ number of degenerate zero-mode solutions,
\begin{align}
\psi_{\pm,0}^{j,|M|}(z,\tau)
&= \left(\frac{|M|}{{\cal A}^2}\right)^{1/4} e^{i\pi |M|(z+\zeta) \frac{{\rm Im}(z+\zeta)}{{\rm Im}\tau}} \sum_{l \in \mathbf{Z}} e^{i\pi |M|\tau \left( \frac{j}{|M|}+l \right)^2} e^{2\pi i|M|(z+\zeta) \left( \frac{j}{|M|}+l \right)} \label{psizero} \\
&= \left(\frac{|M|}{{\cal A}^2}\right)^{1/4} e^{i\pi |M|(z+\zeta) \frac{{\rm Im}(z+\zeta)}{{\rm Im}\tau}}
\vartheta
\begin{bmatrix}
\frac{j}{|M|}\\
0
\end{bmatrix}
(|M|z, |M|\tau), 
\notag
\end{align}
where $j = 0,1,...,|M|-1$ and ${\cal A} = |e_1|^2 {\rm Im}\tau$ is the area of $T^2$. The $\vartheta$ function is defined by
\begin{align}
\vartheta
\begin{bmatrix}
a\\
b
\end{bmatrix}
(\nu, \tau)
=
\sum_{l\in \mathbb{Z}}
e^{\pi i (a+l)^2\tau}
e^{2\pi i (a+l)(\nu+b)},
\end{align}
so-called the Jacobi theta function.

We can  show that Eq.~(\ref{psizero}) satisfies \cite{Abe:2008fi}
\begin{align}
\psi_0^{j,|M|}(-z, \tau) &= \psi_0^{|M|-j,|M|}(z, \tau), \label{-z} 
\end{align}
as well as 
\begin{align}
\psi_0^{j,|M|}(z,\tau) &= \psi_0^{j+|M|,|M|}(z,\tau), \label{mod|M|}
\end{align}
and, later we will use these relations. 

We can extend the above U(1) theory to U(N) super Yang-Mills theory.
We introduce magnetic fluxes along the diagonal direction of U(N), diag$(M,M',\cdots)$.
Then, our theory has several zero-modes, whose Dirac equations include various magnetic fluxes, 
although their zero-mode wavefunctions are written by the above wavefunctions with 
corresponding magnetic fluxes.
The product of such wavefuncitons, $\psi_{\pm,0}^{j,|M|}(z,\tau)\psi_{\pm,0}^{k,|M'|}(z,\tau)$ can be 
expanded by the wavefunctions $\psi_{\pm,0}^{\ell,|M|+|M|'}(z,\tau)$~\cite{Cremades:2004wa}\footnote{
See also Ref.~\cite{Kobayashi:2019fma}.},
\begin{align}
\psi_{\pm,0}^{j,|M|}(z,\tau)~\psi_{\pm,0}^{k,|M'|}(z,\tau) = \sum_{\ell}C^{jk\ell}_{T^2}(\tau)~\psi_{\pm,0}^{\ell,|M|+|M|'}(z,\tau),
\label{eq:product-wf}
\end{align}
where $\ell = j + k +|M|m$ with integer $m$.
The $\tau$-dependent coefficient $C^{jk\ell}_{T^2}(\tau)$ is written explicitly by 
\begin{align}
C^{jk\ell}_{T^2}(\tau) = \vartheta
\begin{bmatrix}
\frac{|M'|j-|M|k + |M||M'|m}{|M||M'|(|M|+|M'|)}\\
0
\end{bmatrix}
(0, |M||M'|(|M|+|M'|)\tau), 
\notag
\end{align}
up to a constant factor.
The coefficient provides us with three-point couplings because 
their couplings are obtained by wavefunction integrals in the compact space, 
\begin{align}
y^{jk\ell}_{T^2} = \int_{T^2} d^2z \psi_{\pm,0}^{j,|M|}(z,\tau)~\psi_{\pm,0}^{k,|M'|}(z,\tau) (\psi_{\pm,0}^{\ell,|M|+|M|'}(z,\tau))^*.
\end{align}
Similarly, $n$-point couplings, 
\begin{align}
y_{j_1,j_2\cdots,j_n} = \int_{T^2} d^2z \psi_{\pm,0}^{j_1,|M^{(1)}|}(z,\tau)~\psi_{\pm,0}^{j_2,|M^{(2)}|}(z,\tau)
\cdots  (\psi_{\pm,0}^{j_n,|M^{(n)}|}(z,\tau))^*,
\end{align}
are also written by products of  $C^{jk\ell}_{T^2}(\tau)$~\cite{Abe:2009dr}.


\section{Modular symmetry in magnetized $T^2$}
\label{SymT2}

In this section, we review the modular symmetry of the zero-mode wavefunctions on the magnetized $T^2$ and its orbifolding by $\mathbb{Z}_N$ twist and shift~\cite{Kikuchi:2020frp}. 
To simplify our analysis, we consider the torus with the magnetic $U(1)$ flux and no Wilson lines. 
However, we can extend this analysis to the models with any flux and non-vanishing Wilson lines without any difficulty.


\subsection{$T^2$ models}
\label{T2models}

First, we briefly review modular transformation of zero-mode wavefunctions on $T^2$. 
(See for the modular symmetry, e.g., \cite{Gunning:1962,Schoeneberg:1974,Koblitz:1984,Bruinier:2008}. )
The torus $T^2$ is constructed by $\mathbb{C}/\Lambda$, where $\Lambda$ is spanned by the basis $e_1,\ e_2$ and characterized by the modulus parameter $\tau = e_2/e_1\ ({\rm Im}\tau > 0)$. 
Then, the same lattice with different modulus parameter is given by the following basis,
\begin{align}
\begin{pmatrix}
e'_2 \\ e'_1
\end{pmatrix}
=
\begin{pmatrix}
a & b \\
c & d
\end{pmatrix}
\begin{pmatrix}
e_2 \\ e_1
\end{pmatrix},
\quad
\gamma =
\begin{pmatrix}
a & b \\
c & d
\end{pmatrix}
\in SL(2,\mathbb{Z}) \equiv \Gamma.
\label{SL2Z}
\end{align}
This $SL(2,\mathbb{Z})$ transformation is generated by two generators,
\begin{align}
S=
\begin{pmatrix}
0 & 1 \\
-1 & 0
\end{pmatrix},
\quad
T=
\begin{pmatrix}
1 & 1 \\
0 & 1
\end{pmatrix},
\label{SandT}
\end{align}
which satisfy the algebra:
\begin{align}
S^2 = -\mathbb{I} \equiv Z,\ S^4 = (ST)^3 = \mathbb{I} = Z^2.
\end{align}
This gives following transformations:
\begin{align}
\gamma: z \equiv \frac{u}{e_1} \rightarrow z' \equiv \frac{u}{e'_1} = \frac{z}{c\tau+d}, \label{zSL2Z} \\
\gamma: \tau \equiv \frac{e_2}{e_1} \rightarrow \tau' \equiv \frac{e'_2}{e'_1} = \frac{a\tau+b}{c\tau+d}, \label{tauSL2Z}
\end{align}
where $u$ is the complex coordinate of $\mathbb{C}$ and $z$ is that of $T^2$. 
Note that the Wilson line $\zeta$ is transformed as in $z$. These are also generated by two generators $S$ and $T$,
\begin{align}
S: (z,\tau) \rightarrow \left( -\frac{z}{\tau}, -\frac{1}{\tau} \right), \quad T: (z,\tau) \rightarrow (z,\tau+1). \label{ztauSandT}
\end{align}
Since $Z=-\mathbb{I}$ leaves $\tau$ invariant, $Z(z,\tau)=(-z,\tau)$, the transformation group for $\tau$ is isomorphic to $\bar{\Gamma} \equiv \Gamma/\{\pm \mathbb{I}\}$. 
Here, we introduce the principal congruence subgroup of level $N$ defined by
\begin{align}
\Gamma(N) \equiv \left\{ h =
\begin{pmatrix}
a' & b' \\
c' & d'
\end{pmatrix} \in \Gamma \biggl|
\begin{pmatrix}
a' & b' \\
c' & d'
\end{pmatrix}
\equiv
\begin{pmatrix}
1 & 0 \\
0 & 1
\end{pmatrix}
\ ({\rm mod}\ N) \right\}. \label{GammaN}
\end{align}
This is the normal subgroup of $\Gamma$, e.g., $\Gamma(1) \simeq \Gamma$. 
Its quotient group is given by
\begin{align}
\Gamma^{\prime}_N \equiv \Gamma/\Gamma(N) = \langle S, T | S^4 = (ST)^3 = T^N = \mathbb{I},\ S^2T = TS^2 \rangle.
\end{align}
Similarly, we can also introduce $\bar{\Gamma}(N) \equiv \Gamma(N)/\{\pm \mathbb{I}\}$ and obtain its quotient group as follows:
\begin{align}
\Gamma_N \equiv \bar{\Gamma}/\bar{\Gamma}(N) = \langle S, T | S^2 = (ST)^3 = T^N = \mathbb{I} \rangle.
\end{align}
The quotient $\Gamma_N$ is isomorphic to $\Gamma_2 \simeq S_3$, $\Gamma_4 \simeq A_4$, $\Gamma_4 \simeq S_4$, and $\Gamma_5 \simeq A_5$.
In addition, $\Gamma^{\prime}_N$ is the double covering group of $\Gamma_N$.
(See e.g., \cite{Liu:2019khw,Novichkov:2020eep,Liu:2020akv}.)

We are now ready to construct the holomorphic functions of $\tau$, the modular forms $f(\tau)$ of integer weight $k$ for $\Gamma(N)$. 
First, we define the automorphy factor $J_k (\gamma, \tau)$ as
\begin{align}
J_k (\gamma, \tau) = (c\tau + d)^k,\quad \gamma =
\begin{pmatrix}
a & b \\
c & d
\end{pmatrix}
\in \Gamma.
\end{align}
It is straightforward to show it satisfies
\begin{align}
J_k(\gamma_1\gamma_2, \tau) = J_k(\gamma_1, \gamma_2(\tau)) J_k(\gamma_2, \tau), \quad \gamma_1, \gamma_2 \in  \Gamma. \label{J}
\end{align}
Then, the modular forms $f(\tau)$ are defined as the functions satisfying the following relation:
\begin{align}
f(\gamma (\tau)) = J_k(\gamma, \tau) \rho(\gamma) f(\tau),\quad \gamma =
\begin{pmatrix}
a & b \\
c & d
\end{pmatrix}
\in \Gamma, 
\end{align}
where 
\begin{align}
\rho(\gamma_2 \gamma_1) &= \rho(\gamma_2) \rho(\gamma_1), \quad \gamma_1, \gamma_2 \in \Gamma, \\
\rho(h) &= \mathbb{I},\quad h \in \Gamma(N),
\end{align}
and therefore $\rho$ is a unitary representation of the quotient group $\Gamma^{\prime}_N = \Gamma/\Gamma(N)$. Since $f(Z(\tau)) = f(\tau)$, there is the constraint, $(-1)^k \rho (Z) = \mathbb{I}$. 
Thus, if $k = {\rm even}$, $\rho$ becomes a representation of $\Gamma_N \equiv \bar{\Gamma}/\bar{\Gamma}(N)$. 
Here, we can extend the modular forms to the half integer weight $k/2$.
(See e.g., \cite{Koblitz:1984,shimura,Duncan:2018wbw}.)
We define the double covering group of $\Gamma = SL(2,\mathbb{Z})$, $\widetilde{\Gamma} \equiv \widetilde{SL} (2,\mathbb{Z})$ as
\begin{align}
\widetilde{\Gamma} \equiv \left\{ [\gamma, \epsilon] \bigl| \gamma \in \Gamma, \ \epsilon \in \{ \pm 1 \} \right\}. \label{Gammatilde}
\end{align}
This $\widetilde{SL}(2,\mathbb{Z})$ group is generated by two generators,
\begin{align}
\widetilde{S} \equiv [S,1],\ \widetilde{T} \equiv [T,1],
\end{align}
which satisfy the algebra:
\begin{align}
\widetilde{S}^2 = [-\mathbb{I},1] \equiv \widetilde{Z}, \ \widetilde{S}^4 = (\widetilde{S} \widetilde{T})^3 = [\mathbb{I},-1] = \widetilde{Z}^2, \ \widetilde{S}^8 = (\widetilde{S} \widetilde{T})^6 = [\mathbb{I},1] \equiv \mathbb{I} = \widetilde{Z}^4, \ \widetilde{Z} \widetilde{T} = \widetilde{T} \widetilde{Z}.
\label{algebradouble}
\end{align}
The normal subgroup of $\widetilde{\Gamma}$, $\widetilde{\Gamma}(N)$ corresponding to $\Gamma(N)$ of $\Gamma$ is defined by
\begin{align}
\widetilde{\Gamma}(N) \equiv \{ [h, \epsilon] \in \widetilde{\Gamma} | h \in \Gamma(N), \epsilon=1 \}. \label{gammatilde2M}
\end{align}
Then, the new automorphy factor $\widetilde{J}_{k/2} (\widetilde{\gamma}, \tau)$ is given by
\begin{align}
\widetilde{J}_{k/2}(\widetilde{\gamma}, \tau) \equiv \epsilon^{k} J_{k/2}(\gamma,\tau) = \epsilon^{k} (c\tau+d)^{k/2}, \quad k \in \mathbb{Z}, \quad
\widetilde{\gamma} = 
\left[\gamma=
\begin{pmatrix}
a & b \\
c & d
\end{pmatrix}, \epsilon \right] \in \tilde{\Gamma}, \label{Jtilde}
\end{align}
where we take $(-1)^{k/2} = e^{-i\pi k/2}$. 
In this extension, the modular forms $\widetilde{f} (\tau)$ of half integer weight $k/2$ are defined as follows,
\begin{align}
\widetilde{f}(\widetilde{\gamma} (\tau)) = \widetilde{J}_{k/2} (\widetilde{\gamma}, \tau) \rho (\widetilde{\gamma})\widetilde{f} (\tau),\quad \widetilde{\gamma} \in \widetilde{\Gamma}, 
\label{modularformk/2}
\end{align}
where $\rho(h) = \mathbb{I}, \ h \in \widetilde{\Gamma}(N)$, that is, $\rho$ is a unitary representation of the quotient group $\widetilde{\Gamma}'_N \equiv \widetilde{\Gamma}/\widetilde{\Gamma}(N)$. 
The algebra of $\widetilde{\Gamma}'_N$ is given by Eq.~(\ref{algebradouble}) added the further relation $\widetilde{T}^N = \mathbb{I}$.

Next, we consider the modular transformation of zero-modes. 
Under $S$ and $T$ transformations in Eq.~(\ref{ztauSandT}), the equation of motion for 2D Weyl spinor $\psi(z)$, Eq.~(\ref{2D massive Dirac}), is preserved. 
The boundary conditions for $\psi(z)$, Eqs.~(\ref{psiz1}) and (\ref{psiztau}), however, are not preserved under $T$ transformation unless $M = {\rm even}$. 
Here and hereafter, we treat only $M = {\rm even}$ case. 
Under $S$ and $T$, the zero-modes in Eq.~(\ref{psizero}) are transformed  as
\begin{align}
&S: \psi_0^{j,|M|}(z, \tau) \rightarrow \psi_0^{j,|M|}\left( -\frac{z}{\tau}, -\frac{1}{\tau} \right) = (-\tau)^{1/2} \sum_{k=0}^{|M|-1} e^{i\pi /4} \frac{1}{\sqrt{|M|}} e^{2\pi i \frac{jk}{|M|}} \psi_0^{k,|M|}(z, \tau), \label{psiS} \\
&T: \psi_0^{j,|M|}(z, \tau) \rightarrow \psi_0^{j,|M|}(z, \tau+1) \ = e^{i\pi \frac{j^2}{|M|}} \psi_0^{j,|M|}(z, \tau). \label{psiT}
\end{align}
By using the modular forms of half integer weight in Eq.~(\ref{modularformk/2}), we can rewrite them as
\begin{align}
\psi_0^{j,|M|}(\widetilde{\gamma}(z,\tau)) &= \widetilde{J}_{1/2}(\widetilde{\gamma}, \tau) \sum_{k=0}^{|M|-1} \rho(\widetilde{\gamma})_{jk} \psi_0^{k,|M|}(z,\tau), \quad \widetilde{\gamma} \in \widetilde{\Gamma}, \label{wavemodularform} \\
\rho(\widetilde{S})_{jk} &= e^{i\pi/4} \frac{1}{\sqrt{|M|}} e^{2\pi i\frac{jk}{|M|}}, \quad
\rho(\widetilde{T})_{jk} = e^{i\pi \frac{j^2}{|M|}} \delta_{j,k}, \label{rhoSandT}
\end{align}
where $\rho(\widetilde{\gamma})$ is a unitary representation of the quotient group $\widetilde{\Gamma}'_{2|M|} \equiv \widetilde{\Gamma}/\widetilde{\Gamma}(2|M|)$:
\begin{align}
\begin{array}{c}
\rho(\widetilde{S})^2 = \rho(\widetilde{Z}),\ \rho(\widetilde{S})^4 = [\rho(\widetilde{S})\rho(\widetilde{T})]^3 = -\mathbb{I}, \ 
\rho(\widetilde{Z})\rho(\widetilde{T}) = \rho(\widetilde{T})\rho(\widetilde{Z}),\ \rho(\widetilde{T})^{2|M|} = \mathbb{I}.
\label{AlgebraonT2}
\end{array}
\end{align}
Thus, the zero-mode wavefunctions on $T^2$ behave as the modular forms of weight 1/2 for $\widetilde{\Gamma} (2|M|)$.


\subsection{$T^2/\mathbb{Z}_N$ twist orbifold models}
\label{T2tmodels}

Here, we review the modular symmetry for the wavefunctions on the magnetized $T^2/\mathbb{Z}_N$ twist orbifolds. 
The  $T^2/\mathbb{Z}_N$ twist orbifolds are obtained by further identifying the complex coordinate of $T^2$, $z$ with the $\mathbb{Z}_N$ discrete rotated points $\alpha^k_N z$, where
\begin{align}
\alpha^k_N \equiv e^{2\pi ik/N}; ^{\forall} k \in \mathbb{Z}_N=\{0,1,2,...,N-1\}.
\end{align}
In this identification, the wavefunctions on the magnetized $T^2/\mathbb{Z}_N$ twist orbifolds, $\psi^{j,|M|}_{T^2/\mathbb{Z}_N^m}(z)$, are required to satisfy the following boundary condition,
\begin{align}
\psi_{T^2/\mathbb{Z}^m_N}^{j,|M|}(\alpha_N z) = \alpha^m_N \psi_{T^2/\mathbb{Z}^m_N}^{j,|M|}(z), \quad m \in \mathbb{Z}_N, \label{shiftT2ZNbase}
\end{align}
and, therefore, can be expressed by liner combinations of the wavefunctions on $T^2$ as~\cite{Abe:2008fi,Abe:2013bca,Abe:2014noa,Kobayashi:2017dyu}
\begin{align}
\psi_{T^2/\mathbb{Z}^m_N}^{j,|M|}(z) &= {\cal N}_N^t \sum_{k=0}^{N-1} (\alpha^m_N)^{-k} \psi_{T^2}^{j,|M|}(\alpha^k_N z),
\end{align}
where ${\cal N}_N^t$ is the normalization factor.
There exist only four consistent orbifolds such as $N = 2,3,4,6$. 
However, except for $N = 2$, the modulus $\tau$  must be fixed to be a certain value for $N = 3,4,6$. 
Thus, any value of $\tau$ is allowed for $N = 2$, that is the full modular symmetry remains for only $N = 2$. 
Now, we focus on the $T^2/\mathbb{Z}_2$ twist orbifold although we can also consider others with the broken modular symmetry.

The zero-mode wavefunctions on $T^2/\mathbb{Z}_2$ twist orbifold, $\psi_{T^2/\mathbb{Z}^m_2}^{j,|M|}(\alpha_2 z,\tau)$, are obtained as follows. 
By Eq.~(\ref{-z}), zero-mode wavefunctions on $T^2$, $\psi^{j,|M|}_0$, satisfy the following relation,
\begin{align}
\psi_0^{j,|M|}(\alpha^m_2 z,\tau) = \psi_0^{j,|M|}((-1)^m z,\tau) = \psi_0^{|M|-j,|M|}(z,\tau), \quad m = 1.
\end{align}
Thus, using Eq.~(\ref{mod|M|}), we can write $\psi_{T^2/\mathbb{Z}^m_2}^{j,|M|}(z,\tau)$ as
\begin{align}
\psi_{T^2/\mathbb{Z}^m_2}^{j,|M|}(z,\tau) = {\cal N}_2^t \left(\psi_0^{j,|M|}(z,\tau) + (-1)^m \psi_0^{|M|-j,|M|}(z,\tau)\right), \quad
{\cal N}_2^t = \left\{
\begin{array}{l}
1/2\quad (j=0,|M|/2)\\
1/\sqrt{2}\quad ({\rm otherwise})
\end{array}
\right. ,
\label{zerotwistonT2}
\end{align}
where $j = 0,1,...,|M|/2$ and $m = 0,1$. 
There are the ($|M|/2+1$) number of $\mathbb{Z}_2$-even modes ($m=0$) and ($|M|/2-1$) -odd modes ($m=1$). 
Under these liner combinations, the formula of modular forms in Eq.~(\ref{wavemodularform}) becomes
\begin{align}
\psi_{T^2/\mathbb{Z}^m_2}^{j,|M|}(z,\tau) &= \widetilde{J}_{1/2}(\widetilde{\gamma}, \tau) \sum_{k=0}^{|M|/2}
\rho_{T^2/\mathbb{Z}_2^m}(\widetilde{\gamma})_{jk} \psi_{T^2/\mathbb{Z}^m_2}^{k,|M|}(z,\tau).
\end{align}
The unitary representation $\rho_{T^2/\mathbb{Z}_2^m}$ is given by
\begin{align}
\rho_{T^2/\mathbb{Z}_2^{0}}(\widetilde{S})_{jk} &= e^{i\pi/4} \frac{2}{\sqrt{|M|}} \cos \left(\frac{2\pi jk}{|M|}\right), \quad
\rho_{T^2/\mathbb{Z}_2^{0}}(\widetilde{T})_{jk} = e^{i\pi \frac{j^2}{|M|}} \delta_{j,k}, \label{rho+SandT} \\
\rho_{T^2/\mathbb{Z}_2^{1}}(\widetilde{S})_{jk} &= e^{i\pi/4} \frac{2i}{\sqrt{|M|}} \sin \left(\frac{2\pi jk}{|M|}\right), \quad
\rho_{T^2/\mathbb{Z}_2^{1}}(\widetilde{T})_{jk} = e^{i\pi \frac{j^2}{|M|}} \delta_{j,k}, \label{rho+SandT}
\end{align}
where $\rho_{T^2/\mathbb{Z}_2^{0}}(\widetilde{S})$ is multiplied by a further factor $1/\sqrt{2}$ for $j$ or $k$ = $0,|M|/2$. 
We can directly check that they satisfy the algebra of $\widetilde{\Gamma}'_{2|M|}$, and the further algebraic relation,
\begin{align}
\begin{array}{c}
\rho_{T^2/\mathbb{Z}_2^m}(\widetilde{S})^2 = \rho_{T^2/\mathbb{Z}_2^m}(\widetilde{Z}) = i(-1)^m.
\end{array}
\label{Algebraontwist}
\end{align}
Thus, the representations on the $T^2/\mathbb{Z}_2$ twist orbifold satisfy the same algebra with $T^2$. 
Note that we have not necessarily obtained the irreducible representation of $\widetilde{\Gamma}'_{2|M|}$. 
Actually, we will see the further decomposition in the end of this section.


\subsection{$T^2/\mathbb{Z}_N$ shift orbifold models}
\label{T2smodels}

As another example, we now review the magnetized $T^2/\mathbb{Z}_N$ shift orbifolds.
The $T^2/\mathbb{Z}_N$ shift orbifolds are obtained by further identifying the complex coordinate of $T^2$, $z$ with the $\mathbb{Z}_N$ discrete shift points $z + ke^{(m,n)}_N$~\cite{Fujimoto:2013xha}, where
\begin{align}
ke^{(m,n)}_N \equiv (m + n\tau)/N; ^{\forall} k, ^{\exists} m, ^{\exists} n \in \mathbb{Z}_N=\{0,1,2,...,N-1\}.
\label{boundaryconditiononshiftT2}
\end{align}
Since the full modular symmetry remains only on the $T^2/\mathbb{Z}_N$ shift orbifolds obtained by further identifying $z$ with $z + ke_N^{(m,n)}$ for all $m,n \in \mathbb{Z}_N$, we consider such full shift orbifolds.

Any $\mathbb{Z}_N$ shift is generated by two shifts, $e_N^{(1,0)} = 1/N$ and $e_N^{(0,1)} = \tau/N$. 
We should consider the identifications with these shifts. 
In these identifications, the wavefunctions on the magnetized $T^2/\mathbb{Z}_N$ shift orbifolds, $\psi_{T^2/\mathbb{Z}^{(\ell_1,\ell_2)}_N}^{j,|M|}(z)$, are required to satisfy the following further boundary conditions,
\begin{align}
\psi_{T^2/\mathbb{Z}^{(\ell_1,\ell_2)}_N}^{j,|M|}(z+e_N^{(1,0)}) &= \alpha^{\ell_1}_N e^{i\chi_N^{(1,0)}(z)} \psi_{T^2/\mathbb{Z}^{(\ell_1,\ell_2)}_N}^{j,|M|}(z), \label{psiz10} \\
\psi_{T^2/\mathbb{Z}^{(\ell_1,\ell_2)}_N}^{j,|M|}(z+e_N^{(0,1)}) &= \alpha^{\ell_2}_N e^{i\chi_N^{(0,1)}(z)} \psi_{T^2/\mathbb{Z}^{(\ell_1,\ell_2)}_N}^{j,|M|}(z), \label{psiz01}
\end{align}
with the $\mathbb{Z}_N$ phase $\alpha_N^{\ell} = e^{2\pi i\ell/N},\ \ell \in \mathbb{Z}_N$ and 
\begin{align}
\chi_N^{(m,n)}(z) = \pi |M| \left(\frac{{\rm Im}(\bar{e}_N^{(m,n)}(z + \tau))}{{\rm Im}\tau} +\frac{mn}{N} \right).
\end{align}
The exponential factor $e^{i\chi_N^{(m,n)}(z)}$ is required to be consistent with the torus boundary conditions in Eqs.~(\ref{psiz1}) and (\ref{psiztau}). 
Moreover, to generate any shift from these two shifts, it should be satisfied that
\begin{align}
\alpha_N^{\ell_1+\ell_2} e^{i\chi_N^{(1,0)}(z+e_N^{(0,1)}) +i\chi_N^{(0,1)}(z)}
 = \alpha_N^{\ell_1+\ell_2} e^{i\chi_N^{(0,1)}(z+e_N^{(1,0)})+i\chi_N^{(1,0)}(z)}
 = \alpha_N^{\ell_{1+2}} e^{i\chi_N^{(1,1)}(z)}.
\end{align}
This shows the conditions for both the magnetic flux and the $\mathbb{Z}_N$ phase as follows:
\begin{align}
\left\{
\begin{array}{l}
M/N^2 \equiv s \in \mathbb{Z},\ \ell_{1+2} = \ell_1 + \ell_2\ ({\rm mod}\ N) \quad ({\rm for}\ N \in \mathbb{Z}) \\
M/N^2 \equiv s \in 2\mathbb{Z}+1,\ \ell_{1+2} = \ell_1 + \ell_2 + N/2\ ({\rm mod}\ N) \quad ({\rm for}\ N \in 2\mathbb{Z}) 
\end{array}
\right. . 
\label{shiftcondition}
\end{align}
Remembering the assumption $M \in 2\mathbb{Z}$, the case of $s \in 2\mathbb{Z}+1,\ N \in 2\mathbb{Z}+1$ is rejected from the above.
Taking these into account , the boundary condition for any $\mathbb{Z}_N$ shift is induced as 
\begin{align}
\psi_{T^2/\mathbb{Z}^{\ell}_N}^{j,|M|}(z + ke_N^{(m,n)}) = (\alpha^{\ell}_N)^k e^{ik\chi_N^{(m,n)}(z)} \psi_{T^2/\mathbb{Z}^{\ell}_N}^{j,|M|}(z),\quad \ell = m\ell_1 + n\ell_2\ ({\rm mod}\ N). \label{twistT2ZNbase}
\end{align}
According to Eq.~(\ref{shiftcondition}), for $s \in 2\mathbb{Z}+1,\ N \in 2\mathbb{Z}$, the extra factor $mnN/2$ is added to $\ell$.
Then, the consistency of the contractible loops on $T^2$ gives the further magnetic flux condition $M/N \equiv t \in \mathbb{Z}$, but it has been already satisfied. 
Thus, the eigenfunctions for  $^{\exists}e_N^{(m,n)}$-shift can be expressed by liner combinations of the wavefunctions on $T^2$ as 
\begin{align}
\psi_{T^2/\mathbb{Z}^{\ell}_N}^{j,|M|}(z) &= {\cal N}_N^s \sum_{k=0}^{N-1} (\alpha^{\ell}_N)^{-k} e^{-ik\chi_N^{(m,n)}(z)}
\psi_{T^2}^{j,|M|}(z + ke_N^{(m,n)}),
\end{align}
where ${\cal N}_N^s$ is the normalization factor. 
Since $e^{-ik\chi_N^{(m,n)}(z)} \psi_{T^2}^{j,|M|}(z + ke_N^{(m,n)})$ satisfies the same equation of motion and boundary conditions with $\psi_{T^2}^{j,|M|}(z)$, it is shown that
\begin{align}
\psi_{T^2}^{j,|M|}(z + ke_N^{(m,n)}) = e^{ik\chi_N^{(m,n)}(z)} e^{i\pi km(2j-(N-k)nN|s|)/N} \psi_{T^2}^{j+knN|s|,|M|}(z).
\end{align}
Eventually, we obtain the eigenfunctions for $^\exists e_N^{(m,n)}$-shift as
\begin{align}
\psi_{T^2/\mathbb{Z}^\ell_N}^{j,|M|}(z,\tau) = {\cal N}_N^s \sum_{k=0}^{N-1} e^{-2\pi ik(\ell-mj)/N} e^{-i\pi k(N-k)mn|s|} \psi_{T^2}^{j+knN|s|,|M|}(z,\tau), \label{shiftT2ZNexp}
\end{align}
where $j=0,...,N|s|-1$. 
To construct the eigenfunctions for $^\forall e_N^{(m,n)}$-shifts, we have to consider the simultaneous eigenfunctions for both $e_N^{(1,0)}$ and $e_N^{(0,1)}$-shifts under the conditions in Eqs.~(\ref{psiz10}), (\ref{psiz01}), and (\ref{shiftcondition}). 
Since the boundary condition for $e_N^{(1,0)}$, Eq.~(\ref{psiz10}), gives the constraint $\ell_1=j$ (mod $N$), we can obtain the eigenfunctions for $^{\forall} e^{(m,n)}_N$-shifts as follows:
\begin{align}
\Psi_{T^2/\mathbb{Z}^{(\ell_1,\ell_2)}_N}^{r,|s|}(z,\tau) &\equiv \psi_{T^2/\mathbb{Z}^{(\ell_1,\ell_2)}_N}^{j,|M|}(z,\tau) = \frac{1}{\sqrt{N}} \sum_{k=0}^{N-1} e^{-2\pi ik\ell_2/N} \psi_{T^2}^{j+kN|s|,|M|}(z,\tau), \label{shiftT2ZNexpall} \\
j&=Nr+\ell_1 \in \mathbb{Z}_{N|s|}, \ r \in \mathbb{Z}_{|s|}, \ \ell_1, \ell_2 \in \mathbb{Z}_N, \notag
\end{align}
where $M$ and $s$ can only take the values allowed in Eq.~(\ref{shiftcondition}). 
There are $|s|$ number of the $\mathbb{Z}_N$ shifts invariant modes ($\ell_1=\ell_2=0$) and ($|M|-|s|$) not invariant modes. 
Under these liner combinations, the formula of modular forms in Eq.~(\ref{wavemodularform}) becomes
\begin{align}
&\Psi_{T^2/\mathbb{Z}^{(\ell_1,\ell_2)}_N}^{r,|s|}(\widetilde{\gamma}(z,\tau)) = \widetilde{J}_{1/2}(\widetilde{\gamma}, \tau) \sum_{r'=0}^{|s|-1} \sum_{\ell'_1,\ell'_2=0}^{N-1} \rho_{T^2/\mathbb{Z}_N^{(\ell_1,\ell_2)}}(\widetilde{\gamma})_{rr',(\ell_1,\ell_2)(\ell'_1,\ell'_2)} \Psi_{T^2/\mathbb{Z}^{(\ell'_1,\ell'_2)}_N}^{r',|s|}(z,\tau),\label{reponT2s}
\end{align}
for $ \widetilde{\gamma} \in \widetilde{\Gamma}$, and unitary matrices are represented by 
\begin{align} 
&\rho_{T^2/\mathbb{Z}_N^{(\ell_1,\ell_2)}}(\widetilde{S})_{rr',(\ell_1,\ell_2)(\ell'_1,\ell'_2)} = e^{i\pi/4} \frac{1}{\sqrt{|s|}} e^{2\pi i \left(\frac{\ell_1}{N}+r\right) \left(\frac{\ell'_1}{N}+r'\right)/|s|} \delta_{\ell_2,\ell'_1} \delta_{N-\ell_1,\ell'_2}, \label{rhoshiftZNS} \\
&\rho_{T^2/\mathbb{Z}_N^{(\ell_1,\ell_2)}}(\widetilde{T})_{rr',(\ell_1,\ell_2)(\ell'_1,\ell'_2)} = e^{i\pi \left(\frac{\ell_1}{N}+r\right)^2/|s|} \delta_{r,r'} \delta_{\ell_1,\ell'_1} \delta_{\ell_2-\ell_1,\ell'_2}, 
\label{rhoshiftZNT}
\end{align}
where $\delta_{\ell_2-\ell_1,\ell_2'}$  in $\rho_{T^2/\mathbb{Z}_N^{(\ell_1,\ell_2)}}(\widetilde{T})$ is modified  into $\delta_{\ell_2-\ell_1+N/2,\ell_2'}$ for $s \in 2\mathbb{Z} + 1$, $N  \in 2\mathbb{Z}$.
Then, for $s \in 2\mathbb{Z}$, we can directly show that the $\mathbb{Z}_N$ shifts invariant modes behave as modular forms for $\widetilde{\Gamma}(2|M|/N^2)$ and  variant modes are that for $\widetilde{\Gamma}(2|M|)$. 
Although it may seem these give the same result for $s \in 2\mathbb{Z} + 1$, 
the modular transformation does not close in 
the $\mathbb{Z}_N$ shift invariant modes, but they transform to $\mathbb{Z}_N$ shift variant modes.
Note that the invariant modes correspond to modes on ${T^2}' \simeq \mathbb{C}/{\Lambda}'$, ${\Lambda}' \equiv \Lambda/N$ with the magnetic flux $M/N^2=s$.
Thus, we can understand this from the fact that the wavefunctions on torus with the magnetic flux $M=2\mathbb{Z} + 1$ are not consistent with $T$ transformation $\tau \rightarrow \tau + 1$.

Instead, there are the $|s|$ number of the $\mathbb{Z}_N$ shift $(\ell_1,\ell_2) = (N/2,N/2)$ modes and the other $(|M|-|s|)$ modes transformed independently under the modular transformation. 
Also we can check that the former modes behave as the modular forms for $\widetilde{\Gamma}(8|M|/N^2)$ and they satisfy the further algebraic relation,
\begin{align}
\rho_{T^2/\mathbb{Z}_N^{(N/2,N/2)}}(\widetilde{T})^{2|M|/N^2} = i \mathbb{I}.
\label{shiftodd}
\end{align}
The latter modes just behave as that for $\widetilde{\Gamma}(2|M|)$.


\subsection{$T^2/\mathbb{Z}_N$ twist and shift orbifold models}
\label{T2tsmodels}

As the end of the review of the $T^2/\mathbb{Z}_N$ orbifolds, we study the $T^2/\mathbb{Z}_N$ twist and shift orbifold models. 
The full modular symmetry remains only on the combination of the $T^2/\mathbb{Z}_2$ twist orbifold and the full $T^2/\mathbb{Z}_2$ shift orbifold, that is the full $T^2/\mathbb{Z}_2$ twist and shift orbifold,
since the full $T^2/\mathbb{Z}_2$ shift orbifold only satisfies the consistency condition with the $T^2/\mathbb{Z}_2$ twist orbifold,~i.e., $N-\ell_{1,2} \equiv \ell_{1,2}\ ({\rm mod}\ N)$.
For $M/4=s \in 2\mathbb{Z}$, the wavefunctions on the above orbifold are given by
\begin{align}
&\Psi_{T^2/\mathbb{Z}^{(m;\ell_1,\ell_2)}_2}^{r,|s|}
= {\cal N}_2^{st}  \left( \Psi_{T^2/\mathbb{Z}^{(\ell_1,\ell_2)}_2}^{r,|s|} + (-1)^{m+\ell_2} \Psi_{T^2/\mathbb{Z}^{(\ell_1,\ell_2)}_2}^{|s|-r-\ell_1,|s|} \right)
= {\cal N}_2^{st} \left( \psi_{T^2/\mathbb{Z}_2^{m}}^{2r+\ell_1,4|s|} + (-1)^{\ell_2} \psi_{T^2/\mathbb{Z}_2^{m}}^{2r+\ell_1+2|s|,4|s|} \right) \notag \\
&= {\cal N}_2^{st} \left( \psi_{T^2}^{2r+\ell_1,4|s|} + (-1)^{\ell_2+m} \psi_{T^2}^{2(|s|-r-\ell_1)+\ell_1,4|s|} + (-1)^{\ell_2} \psi_{T^2}^{2(|s|+r)+\ell_1,4|s|} + (-1)^{m} \psi_{T^2}^{2(2|s|-r-\ell_1)+\ell_1,4|s|} \right), \notag \\
&\hspace{5.0cm} s \in 2\mathbb{Z}, \ r \in \mathbb{Z}_{\frac{|s|}{2}+1-\ell_1}, \ m, \ell_1, \ell_2 \in \mathbb{Z}_2, \label{twistshiftT2Z2exp}
\end{align}
where ${\cal N}_2^{st}$ is the normalization factor. 
There are the $(|M|/8+1)$ and $(|M|/8-1)$ numbers of the full $\mathbb{Z}_2$ shifts invariant modes 
$(m;\ell_1,\ell_2) = (0;0,0)$ and $ (1;0,0)$, respectively. 
Under these liner combinations, the unitary representations in Eq.~(\ref{rhoSandT}) become
\begin{align}
\rho_{T^2/\mathbb{Z}_2^{(0;\ell_1,\ell_2)}}(\widetilde{S})_{rr',(\ell_1,\ell_2)(\ell'_1,\ell'_2)} &= e^{i\pi/4} \frac{2}{\sqrt{|s|}}  \cos \left( 2\pi (\ell_1/2+r) (\ell'_1/2+r')/|s| \right) \delta_{\ell_2,\ell'_1} \delta_{\ell_1,\ell'_2}, \label{rho+shiftZ2S} \\
\rho_{T^2/\mathbb{Z}_2^{(0;\ell_1,\ell_2)}}(\widetilde{T})_{rr',(\ell_1,\ell_2)(\ell'_1,\ell'_2)} &= e^{i\pi \left(\frac{\ell_1}{N}+r\right)^2/|s|} \delta_{r,r'} \delta_{\ell_1,\ell'_1} \delta_{\ell_2-\ell_1,\ell'_2}, \label{rho+shoftZ2T} \\
\rho_{T^2/\mathbb{Z}_2^{(1;\ell_1,\ell_2)}}(\widetilde{S})_{rr',(\ell_1,\ell_2)(\ell'_1,\ell'_2)} &= e^{i\pi/4} \frac{2i}{\sqrt{|s|}}  \sin \left( 2\pi (\ell_1/2+r) (\ell'_1/2+r')/|s| \right) \delta_{\ell_2,\ell'_1} \delta_{\ell_1,\ell'_2}, \label{rho-shiftZ2S} \\
\rho_{T^2/\mathbb{Z}_2^{(1;\ell_1,\ell_2)}}(\widetilde{T})_{rr',(\ell_1,\ell_2)(\ell'_1,\ell'_2)} &= e^{i\pi \left(\frac{\ell_1}{N}+r\right)^2/|s|} \delta_{r,r'} \delta_{\ell_1,\ell'_1} \delta_{\ell_2-\ell_1,\ell'_2}. \label{rho-shoftZ2T}
\end{align}
Then, we can directly show that the $\mathbb{Z}_2$ shifts invariant modes $(m;\ell_1,\ell_2) = (m;0,0)$ behave as the modular forms for $\widetilde{\Gamma}(|M|/2)$ and variant modes are that for $\widetilde{\Gamma}(2|M|)$. 
Moreover, they satisfy the further algebraic relation such as Eq.~(\ref{Algebraontwist}).

The same argument is also possible for $s \in 2\mathbb{Z} +1$, but we show only the results here.
There are the $(|M|/4 - 1)$ and $(|M|/4 + 1)$ numbers of the $\mathbb{Z}_2$ twist and full shifts $(m;\ell_1,\ell_2) = (0;1,1)$ and 
$(1,1,1)$ modes, respectively. 
They are transformed independently under the modular transformation.
The $\mathbb{Z}_2$ twist and full shifts $(m;1,1)$ modes behave as the modular forms for $\widetilde{\Gamma}(2|M|)$ and they satisfy the further algebraic relation such as Eqs.~(\ref{Algebraontwist}) and (\ref{shiftodd}).
Other modes just behave as that for $\widetilde{\Gamma}(2|M|)$ and satisfy Eq.~(\ref{Algebraontwist}).


\section{Modular symmetry in magnetized $T^2_1\times T^2_2$}
\label{SymT2xT2}

In the previous section, we have seen the modular symmetry on the magnetized $T^2$ and its orbifolds by the $\mathbb{Z}_N$ twist and shift. 
In this section, let us consider the modular symmetry of the zero-mode wavefunctions on the magnetized $T^2_1 \times T^2_2$ and orbifolds where the complex modulus parameters are identified as $\tau_1 = \tau_2 \equiv \tau$.
As in the previous analyses on $T^2$, we assume the even magnetic fluxes and focus on the zero-mode wavefunctions on the orbifolds.


\subsection{$T^2_1\times T^2_2$ models}
\label{T2xT2models} 

Since the wavefunctions on $T^2$ behave like the modular forms of weight 1/2 for $\widetilde{\Gamma} (2|M|)$, we can treat the wavefunctions on $T^2 _1\times T^2_2$ as the modular forms of weight 1 as follows:
\begin{align}
\psi^{j,|M_1|}_{0,T^2_1}({\gamma}(z_1,\tau)) \psi^{k,|M_2|}_{0,T^2_2}({\gamma}(z_2,\tau))
&= {J}_1({\gamma},\tau) \sum_{m=0}^{|M_1|-1}\rho({\gamma})_{jm} \sum_{n=0}^{|M_2|-1}\rho({\gamma})_{kn} 
\psi^{m,|M_1|}_{0,T^2_1}(z_1,\tau) \psi^{n,|M_2|}_{0,T^2_2}(z_2,\tau) \notag \\
&\equiv {J}_1({\gamma},\tau) \sum_{m=0}^{|M_1|-1} \sum_{n=0}^{|M_2|-1} \rho({\gamma})_{(jk)(mn)} \psi^{m,|M_1|}_{0,T^2_1}(z_1,\tau) \psi^{n,|M_2|}_{0,T^2_2}(z_2,\tau), 
\label{modularformT2xT2} \\
\rho(S)_{(jk)(mn)} = \prod_{t=1,2} \rho_{T^2_t}(\widetilde{S})_{j_tm_t} 
&= \rho_{T^2_1}(\widetilde{S})_{jm} \rho_{T^2_2}(\widetilde{S})_{kn}
= \frac{i}{\sqrt{|M_1M_2|}} e^{2i\pi (\frac{jm}{|M_1|} + \frac{kn}{|M_2|})},\\
\rho(T)_{(jk)(mn)} = \prod_{t=1,2} \rho_{T^2_t}(\widetilde{T})_{j_tm_t} 
&= \rho_{T^2_1}(\widetilde{T})_{jm} \rho_{T^2_2}(\widetilde{T})_{kn}
= e^{i\pi (\frac{j^2}{|M_1|} + \frac{k^2}{|M_2|})} \delta_{j,m} \delta_{k,n}, 
\label{SandTonT2xT2twist}\\
&j,m \in \mathbb{Z}_{|M_1|},k,n \in \mathbb{Z}_{|M_2|}, {\gamma} \in {\Gamma}, \notag
\end{align}
where the lower indices 1 and 2 of the coordinates $z$ and the magnetic fluxes $M$ denote the tori $T^2_1$ and $T^2_2$, respectively. 
Note that the modular symmetry on $T^2_1 \times T^2_2$, $\Gamma \times \Gamma$, is broken to $\Gamma$ by the identification $\tau_1 = \tau_2 \equiv \tau$. 
Similarly, the unitary representation $\rho({\gamma})_{(jk)(mn)}$ is broken from $\widetilde{\Gamma}'_{2|M_1|} \times \widetilde{\Gamma}'_{2|M_2|}$ to its subgroup. 
Since $\rho({\gamma})_{(jk)(mn)}$ is given by the tensor products of the representations on $T^2_1$ and $T^2_2$ such as above, by multiplying both algebraic relations in Eq.~(\ref{AlgebraonT2}) for $T_1^2$ and $T^2_2$, we can obtain the following relations for $T^2_1 \times T^2_2$:
\begin{align}
\begin{array}{c}
\rho(T)^{2{\rm lcm}(|M_1|,|M_2|)}_{(jk)(mn)} = \rho(S)^4_{(jk)(mn)} = [\rho(S)\rho(T)]^3_{(jk)(mn)} = \delta_{(jk),(mn)}, \\
{[\rho(S)^2\rho(T)]}_{(jk)(mn)} = [\rho(T)\rho(S)^2]_{(jk)(mn)},\ \rho(S)^2_{(jk)(mn)} = -\delta_{j,|M_1|-m} \delta_{k,|M_2|-n}.
\end{array}
\end{align}
This is just the algebra of $\Gamma_{2{\rm lcm}(|M_1|,|M_2|)}'$. \footnote{2lcm$(a,b)$ denotes two times the least common multiple of $a$ and $b$.}
Thus, the zero-mode wavefunctions on $T^2_1 \times T^2_2$ behave as the modular forms of weight 1 for $\Gamma(2{\rm lcm}(|M_1|,|M_2|))$. 
This argument on the algebraic relations is valid for other orbifolds unless we study orbifolding across  the tori $T^2_1$ and $T^2_2$. 
Such orbifolding by permutation may affect the algebraic relations since their representations cannot be written as the tensor products. 
The remaining of this section, we consider orbifolding $T^2_1 \times T^2_2$ by the $\mathbb{Z}_2$ twist, the full $\mathbb{Z}_N$ shifts and the $\mathbb{Z}_2$ permutation that interchanges the two tori coordinates, $z_1 \leftrightarrow z_2$. 

Before the end of this subsection, we comment on products of wavefunctions and 
couplings.
Using Eq.~(\ref{eq:product-wf}), we can expand the product of wavefunctions 
on $T^2_1 \times T^2_2$ as
\begin{align}
\psi^{j,k}_{0,T^2\times T^2}(z_1,z_2,\tau)~\psi^{j',k'}_{0,T^2\times T^2}(z_1,z_2,\tau)= 
\sum_{j'', k'''}C^{(j,j',j''),(k,k',k'l')}(\tau) ~\psi^{j'',k''}_{0,T^2\times T^2}(z_1,z_2,\tau),
\label{eq:product-wf-2}
\end{align}
where 
\begin{align}
& 
\psi^{j,k}_{0,T^2\times T^2}(z_1,z_2,\tau) = \psi^{j,|M_1|}_{0,T^2_1}(z_1,\tau) \psi^{k,|M_2|}_{0,T^2_2}(z_2,\tau), \\
& C^{(j,j',j''),(k,k',k'')}(\tau) = C^{jj'j''}_{T^2}(\tau)  C^{kk'k''}_{T^2}(\tau).
\end{align}
The modular transformation behaviors in left and right hand sides in Eq.~(\ref{eq:product-wf-2}) 
must be the same.
The $\tau$-dependent coefficient $C^{(j,j',j''),(k,k',k'')}(\tau) $ is the modular form of weight 1.
In particular, when the magnetic fluxes for $\psi^{j,k}_{0,T^2\times T^2}(z_1,z_2,\tau)$ and 
$\psi^{j',k'}_{0,T^2\times T^2}(z_1,z_2,\tau)$ are the same, 
$C^{(j,j',j''),(k,k',k'')}(\tau) $ are multiplets under $\Gamma_{2{\rm lcm}(|M_1|,|M_2|)}'$.
That is, the three-point couplings are the modular form of weight 1 with a non-trivial representation
of $\Gamma_{2{\rm lcm}(|M_1|,|M_2|)}'$.
The $n$-point couplings are also obtained by products of $C^{(j,j',j''),(k,k',k'')}(\tau) $, 
and they are modular forms of weight $(n-2)$.


\subsection{$(T^2_1\times T^2_2)/\mathbb{Z}_N$ twist and shift orbifold models}
\label{T2xT2tsmodels}

First of all, we consider orbifolding by the $\mathbb{Z}_2$ twist and the full $\mathbb{Z}_N$ shift, where the algebraic relations in the previous section is valid. 
On the $(T^2_1\times T^2_2)/\mathbb{Z}_N$ twist and shift orbifolds, in general, the tensor product of the representations of $\widetilde{\Gamma}_{2|M_1|}'$ and $\widetilde{\Gamma}_{2|M_2|}'$ gives the representation of $\Gamma_{2{\rm lcm}(|M_1|,|M_2|)}'$.

Since the wavefunctions on the above orbifolds are obtained by the tensor products of each orbifold, for example, the wavefunctions on the pair of $T^2_1$ and $T^2_2$ with $M_1 = M_2 = 2$ are obtained as
\begin{align}
\begin{pmatrix}
\psi^{(00)}_{T_1^2 \times T^2_2}(z_1,z_2) \\
\psi^{(10)}_{T_1^2 \times T^2_2}(z_1,z_2) \\
\psi^{(01)}_{T_1^2 \times T^2_2}(z_1,z_2) \\
\psi^{(11)}_{T_1^2 \times T^2_2}(z_1,z_2)
\end{pmatrix}
= 
\begin{pmatrix}
\psi^{0,2}_{0,T^2_1}(z_1) \psi^{0,2}_{0,T^2_2}(z_2) \\
\psi^{1,2}_{0,T^2_1}(z_1) \psi^{0,2}_{0,T^2_2}(z_2) \\
\psi^{0,2}_{0,T^2_1}(z_1) \psi^{1,2}_{0,T^2_2}(z_2) \\
\psi^{1,2}_{0,T^2_1}(z_1) \psi^{1,2}_{0,T^2_2}(z_2) 
\end{pmatrix},
\end{align}
while the $\mathbb{Z}_N$-shift even modes on the pair of $T^2_1$ and the $T^2_2/\mathbb{Z}_N$ shift orbifold with $M_1 = 2,\ M_2 = 2N^2$ are
\begin{align}
\begin{pmatrix}
\psi^{(00)}_{T_1^2 \times (T^2_2/\mathbb{Z}^0_N)}(z_1,z_2) \\
\psi^{(10)}_{T_1^2 \times (T^2_2/\mathbb{Z}^0_N)}(z_1,z_2) \\
\psi^{(01)}_{T_1^2 \times (T^2_2/\mathbb{Z}^0_N)}(z_1,z_2) \\
\psi^{(11)}_{T_1^2 \times (T^2_2/\mathbb{Z}^0_N)}(z_1,z_2)
\end{pmatrix}
= 
\begin{pmatrix}
\psi^{0,2}_{0,T^2_1}(z_1) \Psi^{0,2}_{T^2_2/\mathbb{Z}_N^{(0,0)}}(z_2) \\
\psi^{1,2}_{0,T^2_1}(z_1) \Psi^{0,2}_{T^2_2/\mathbb{Z}_N^{(0,0)}}(z_2) \\
\psi^{0,2}_{0,T^2_1}(z_1) \Psi^{1,2}_{T^2_2/\mathbb{Z}_N^{(0,0)}}(z_2) \\
\psi^{1,2}_{0,T^2_1}(z_1) \Psi^{1,2}_{T^2_2/\mathbb{Z}_N^{(0,0)}}(z_2) 
\end{pmatrix}.
\end{align}
Then, the representations of the $S,\ T$ transformations are same on both wavefunctions,
\begin{align}
\rho(S) = \frac{i}{2}
\begin{pmatrix}
1 & 1 & 1 & 1\\
1 & -1 & 1 & -1\\
1 & 1 & -1 & -1\\
1 & -1 & -1 & 1
\end{pmatrix},\quad
\rho(T) =
\begin{pmatrix}
1 & 0 & 0 & 0\\
0 & i & 0 & 0\\
0 & 0 & i & 0\\
0 & 0 & 0 & -1
\end{pmatrix}.\label{eq:S4'-4}
\end{align}
They generate the group $\Gamma'_4 \simeq S'_4$ which has the order 48.
From the correspondence of the $\mathbb{Z}_N$-shift even modes to the 1/$N$ torus with the magnetic flux $M_2/N^2 = 2$, we can understand these equalities.
Now, it is straightforward to confirm that they satisfy the above general rule for the algebraic relations on the $(T^2_1\times T^2_2)/\mathbb{Z}_N$ twist and shift orbifolds, hence, the product representation of $\widetilde{\Gamma}_{4}'$ and $\widetilde{\Gamma}_{4}'$ gives the representation of $\Gamma_{4}'$.
The matrices $\rho(S)$ and $\rho(T)$ in Eq.~(\ref{eq:S4'-4}) correspond to a reducible representation $\Gamma'_4 \simeq S'_4$.
They can be decomposed into a triplet and a singlet.
The triplet corresponds to 
\begin{align}
\begin{pmatrix}
\psi^{(00)}_{T_1^2 \times (T^2_2/\mathbb{Z}^0_N)}(z_1,z_2) \\
\frac{1}{\sqrt{2}}\left(\psi^{(10)}_{T_1^2 \times (T^2_2/\mathbb{Z}^0_N)}(z_1,z_2) +
\psi^{(01)}_{T_1^2 \times (T^2_2/\mathbb{Z}^0_N)}(z_1,z_2) \right)\\
\psi^{(11)}_{T_1^2 \times (T^2_2/\mathbb{Z}^0_N)}(z_1,z_2), 
\end{pmatrix} ,
\end{align}
where $S$ and $T$ are expressed as follows:
\begin{align}
\rho_ (S) = \frac{i}{2}
\begin{pmatrix}
1 & \sqrt{2} & 1\\
\sqrt{2} & 0 & -\sqrt{2} \\
1 & -\sqrt{2} & 1
\end{pmatrix},\quad
\rho (T) =
\begin{pmatrix}
1 & 0 & 0\\
0 & i & 0 \\
0 & 0 & -1
\end{pmatrix}. \label{eq:S4'-3}
\end{align}
In addition, the singlet corresponds to
\begin{align}
\frac{1}{\sqrt{2}}\left(\psi^{(10)}_{T_1^2 \times (T^2_2/\mathbb{Z}^0_N)}(z_1,z_2) -
\psi^{(01)}_{T_1^2 \times (T^2_2/\mathbb{Z}^0_N)}(z_1,z_2) \right),
\end{align}
where $S$ and $T$ are expressed as 
\begin{align}
\rho(S)=-i,\qquad \rho(T)=i.
\end{align}
The wavefunction of the singlet vanishes at $z_1=z_2=0$, while the other do not vanish.
Thus, the singlet is trivial as the conventional modular form $f(\tau)$.

Similarly, we can study other types of orbifolding.
However, there are exceptions on the pair of the $\mathbb{Z}_2$ twist orbifolds and that of the $\mathbb{Z}_N$ shift orbifolds.
In the former case, since the representations on the $T^2/\mathbb{Z}_2$ twist orbifolds satisfy the relation in Eq.~(\ref{Algebraontwist}), the tensor product of these obeys
\begin{align}
\rho(T)^{2{\rm lcm}(|M_1|,|M_2|)} = \rho(S)^4 &= [\rho(S)\rho(T)]^3 = \mathbb{I},\ \rho(S)^2 = -(-1)^{m_1+m_2} \mathbb{I},
\end{align} 
where $m_1,m_2$ denote the $\mathbb{Z}_2$-twist eigenmodes $0,1$ on $T^2_1$ and $T^2_2$, respectively. 
Therefore, the products of the same $\mathbb{Z}_2$-twist eigenmodes $(m_1 = m_2)$ correspond to $\Gamma_{2{\rm lcm}(|M_1|,|M_2|)}'$, while the different $\mathbb{Z}_2$-twist eigenmodes $(m_1 \neq m_2)$ correspond to $\Gamma_{2{\rm lcm}(|M_1|,|M_2|)}$.

In the latter case, only on the pair of the $\mathbb{Z}_N$ shift orbifolds with the magnetic fluxes $M=N^2s,\ s \in 2\mathbb{Z}+1,\ N \in 2\mathbb{Z}$, there are the further relation in Eq.~(\ref{shiftodd}) for the $\mathbb{Z}_N$-shift $(\ell_1,\ell_2)=(N/2,N/2)$ modes.
Then, the tensor products of the algebraic relations obey
\begin{align}
\rho(T)^{4{\rm lcm}(|s_1|,|s_2|)} = \rho(S)^4 &= [\rho(S)\rho(T)]^3 = \mathbb{I},\ \rho(S)^2\rho(T) = \rho(T)\rho(S)^2.
\end{align} 
This means that the products of the $\mathbb{Z}_N$-shift $(\ell_1,\ell_2)=(N/2,N/2)$ modes correspond to $\Gamma_{4{\rm lcm}(|s_1|,|s_2|)}'$, where 4lcm$(a,b)$ denotes 4 times the least common multiple of $a$ and $b$.

With these in mind, we show the algebraic relations for the unitary representation on the $(T^2_1 \times T^2_2)/\mathbb{Z}_N$ twist and shift orbifolds in Table 1. 
The dimension of each normal subspace (eigenmode) is given by the products of the number of modes on each $T^2/\mathbb{Z}_N$ orbifold discussed in section 3. 
It is shown in Table 2.
For simplicity, we omit the results of the $\mathbb{Z}_N$ shift orbifolds with the magnetic fluxes $M=N^2s,\ s \in 2\mathbb{Z}+1,\ N \in 2\mathbb{Z}$, but we can obtain them from the arguments up to now.

\newpage

\begin{align}
\begingroup
\renewcommand{\arraystretch}{1.35}
\begin{array}{c|c|c|c} \hline
T^2_1 & T^2_2 & {\rm Magnetic\ flux} & {\rm Algebra\ for\ each\ mode} \\ \hline \hline
T^2 & T^2,\  \mathbb{Z}_2\ {\rm twist} & M_1,M_2 & \Gamma_{2{\rm  lcm}(|M_1|,|M_2|)}' \\ \hline
T^2 & 
\begin{array}{c}
\mathbb{Z}_N\ {\rm shift},\\
\mathbb{Z}_2\ {\rm twist\ \&\ shift}
\end{array}
& 
\begin{array}{c}
M_1,M_2=N^2_2s_2\\
(s_2 \in 2\mathbb{Z})
\end{array}
& \left\{ 
\begin{array}{l} 
\Gamma_{2{\rm lcm}(|M_1|,|s_2|)}'(+_2^s)\\ 
\Gamma_{2{\rm lcm}(|M_1|,|M_2|)}'(-_2^s) 
\end{array} \right. \\ \hline
\mathbb{Z}_2\ {\rm twist} & \mathbb{Z}_2\ {\rm twist} & M_1,M_2 & \left\{ 
\begin{array}{l} 
\Gamma_{2{\rm lcm}(|M_1|,|M_2|)}'(\pm_1^t\pm_2^t)\\ 
\Gamma_{2{\rm lcm}(|M_1|,|M_2|)}(\pm_1^t\mp_2^t) 
\end{array} \right. \\ \hline
\mathbb{Z}_2\ {\rm twist} & \mathbb{Z}_N\ {\rm shift} & 
\begin{array}{c}
M_1,M_2=N^2_2s_2 \\
(s_2 \in 2\mathbb{Z})
\end{array}
& \left\{ 
\begin{array}{l} 
\Gamma_{2{\rm lcm}(|M_1|,|s_2|)}'(+_2^s)\\ 
\Gamma_{2{\rm lcm}(|M_1|,|M_2|)}'(-_2^s) 
\end{array} \right. \\ \hline
\mathbb{Z}_2\ {\rm twist} & \mathbb{Z}_2\ {\rm twist\ \&\ shift} & 
\begin{array}{c}
M_1,M_2=4s_2\\
(s_2 \in 2\mathbb{Z})
\end{array}
& \left\{ 
\begin{array}{l} 
\Gamma_{2{\rm lcm}(|M_1|,|s_2|)}'(\pm_1^t\pm_2^t+_2^s)\\ 
\Gamma_{2{\rm lcm}(|M_1|,|s_2|)}(\pm_1^t\mp_2^t+_2^s)\\
\Gamma_{2{\rm lcm}(|M_1|,|M_2|)}'(\pm_1^t\pm_2^t-_2^s)\\ 
\Gamma_{2{\rm lcm}(|M_1|,|M_2|)}(\pm_1^t\mp_2^t-_2^s)
\end{array} \right. \\ \hline
\mathbb{Z}_N\ {\rm shift} & 
\begin{array}{c}
\mathbb{Z}_N\ {\rm shift},\\ 
\mathbb{Z}_2\ {\rm twist\ \&\ shift} 
\end{array}
&
\begin{array}{c}
M_1=N^2_1s_1,\\
M_2=N^2_2s_2\\
(s_1,s_2 \in 2\mathbb{Z})
\end{array}
& \left\{ 
\begin{array}{l} 
\Gamma_{2{\rm lcm}(|s_1|,|s_2|)}'(+_1^s+_2^s)\\ 
\Gamma_{2{\rm lcm}(|s_1|,|M_2|)}'(+_1^s-_2^s)\\
\Gamma_{2{\rm lcm}(|M_1|,|s_2|)}'(-_1^s+_2^s)\\ 
\Gamma_{2{\rm lcm}(|M_1|,|M_2|)}'(-_1^s-_2^s) 
\end{array} \right. \\ \hline
\mathbb{Z}_2\ {\rm twist\ \&\ shift} & \mathbb{Z}_2\ {\rm twist\ \&\ shift} & 
\begin{array}{c}
M_1=4s_1,\\
M_2=4s_2 \\
(s_1,s_2 \in 2\mathbb{Z})
\end{array}
& \left\{ 
\begin{array}{l} 
\Gamma_{2{\rm lcm}(|s_1|,|s_2|)}'(\pm_1^t+_1^s\pm_2^t+_2^s)\\ 
\Gamma_{2{\rm lcm}(|s_1|,|s_2|)}(\pm_1^t+_1^s\mp_2^t+_2^s)\\
\Gamma_{2{\rm lcm}(|s_1|,|M_2|)}'(\pm_1^t+_1^s\pm_2^t-_2^s)\\ 
\Gamma_{2{\rm lcm}(|s_1|,|M_2|)}(\pm_1^t+_1^s\mp_2^t-_2^s)\\
\Gamma_{2{\rm lcm}(|M_1|,|s_2|)}'(\pm_1^t-_1^s\pm_2^t+_2^s)\\ 
\Gamma_{2{\rm lcm}(|M_1|,|s_2|)}(\pm_1^t-_1^s\mp_2^t+_2^s)\\
\Gamma_{2{\rm lcm}(|M_1|,|M_2|)}'(\pm_1^t-_1^s\pm_2^t-_2^s)\\ 
\Gamma_{2{\rm lcm}(|M_1|,|M_2|)}(\pm_1^t-_1^s\mp_2^t-_2^s)
\end{array} \right. \\ \hline
\end{array} \notag
\endgroup
\end{align}
Table 1: The algebraic relations for the unitary representation on the $(T^2_1 \times T^2_2)/\mathbb{Z}_N$ twist and shift orbifolds. 
The first (second) column shows the types of orbifolds from $T^2_1$ ($T^2_2$), which also include $T^2_1$ ($T^2_2$) itself.
The third column shows the flux condition on each orbifold. 
The last column shows the algebraic relations that the unitary representation on the orbifolds at least satisfies. 
The sign $(+/-_{1/2}^{t/s})$ means the invariant $(+)$/variant $(-)$ modes under the $\mathbb{Z}_2$ twist $(^t)$/$\mathbb{Z}_N$ shift $(^s)$ on $T^2_1\ (_1)$/$T^2_2\ (_2)$. 

\newpage
\begin{align}
\begingroup
\renewcommand{\arraystretch}{1.1}
\begin{array}{c|c|c|c} \hline
T^2_1 & T^2_2 & {\rm Normal\ subspaces} & {\rm \#\ of\ zero}$-${\rm modes} \\ \hline \hline
T^2 & T^2 & $-$ & |M_1||M_2| \\ \hline
T^2 & \mathbb{Z}_2\ {\rm twist} &
\begin{array}{c}
\pm_2^t
\end{array}
&
\begin{array}{c}
|M_1|(|M_2|\pm2)/2
\end{array}
\\ \hline
T^2 & \mathbb{Z}_N\ {\rm shift} &
\begin{array}{c}
+_2^s \\
-_2^s
\end{array}
& 
\begin{array}{c}
|M_1||s_2|\\
|M_1||s_2|(N^2_2-1)
\end{array}
\\ \hline
T^2 & \mathbb{Z}_2\ {\rm twist \ \&\ shift} &
\begin{array}{c}
\pm_2^t+_2^s \\
\pm_2^t-_2^s
\end{array}
& 
\begin{array}{c}
|M_1|(|s_2|\pm2)/2\\
3|M_1||s_2|/2
\end{array}
\\ \hline
\mathbb{Z}_2\ {\rm twist} & \mathbb{Z}_2\ {\rm twist} &
\begin{array}{c}
\pm_1^t\pm_2^t \\
\pm_1^t\mp_2^t
\end{array}
& 
\begin{array}{c}
(|M_1|\pm2)(|M_2|\pm2)/4\\
(|M_1|\pm2)(|M_2|\mp2)/4
\end{array}
\\ \hline
\mathbb{Z}_2\ {\rm twist} & \mathbb{Z}_N\ {\rm shift} &
\begin{array}{c}
\pm_1^t+_2^s \\
\pm_1^t-_2^s
\end{array}
& 
\begin{array}{c}
(|M_1|\pm2)|s_2|/2\\
(|M_1|\pm2)|s_2|(N^2_2-1)/2
\end{array}
\\ \hline
\mathbb{Z}_2\ {\rm twist} & \mathbb{Z}_2\ {\rm twist\ \&\ shift} &
\begin{array}{c}
\pm_1^t\pm_2^t+_2^s \\
\pm_1^t\pm_2^t-_2^s \\
\pm_1^t\mp_2^t+_2^s \\
\pm_1^t\mp_2^t-_2^s
\end{array}
& 
\begin{array}{c}
(|M_1|\pm2)(|s_2|\pm2)/8\\
3(|M_1|\pm2)|s_2|/8\\
(|M_1|\pm2)(|s_2|\mp2)/8\\
3(|M_1|\pm2)|s_2|/8
\end{array}
\\ \hline
\mathbb{Z}_N\ {\rm shift} & \mathbb{Z}_N\ {\rm shift} &
\begin{array}{c}
+_1^s+_2^s \\
+_1^s-_2^s \\
-_1^s+_2^s \\
-_1^s-_2^s
\end{array}
& 
\begin{array}{c}
|s_1||s_2|\\
|s_1||s_2|(N^2_2-1)\\
|s_1||s_2|(N^2_1-1)\\
|s_1|s_2|(N^2_1-1)(N^2_2-1)
\end{array}
\\ \hline
\mathbb{Z}_N\ {\rm shift} & \mathbb{Z}_2\ {\rm twist\ \&\ shift} &
\begin{array}{c}
+_1^s\pm_2^t+_2^s \\
+_1^s\pm_2^t-_2^s \\
-_1^s\pm_2^t+_2^s \\
-_1^s\pm_2^t-_2^s
\end{array}
& 
\begin{array}{c}
|s_1|(|s_2|\pm2)/2\\
3|s_1||s_2|/2\\
|s_1|(N^2_1-1)(|s_2|\pm2)/2\\
3|s_1||s_2|(N^2_1-1)/2
\end{array}
\\ \hline
\mathbb{Z}_2\ {\rm twist\ \&\ shift} & \mathbb{Z}_2\ {\rm twist\ \&\ shift} &
\begin{array}{c}
\pm_1^t+_1^s\pm_2^t+_2^s \\
\pm_1^t+_1^s\pm_2^t-_2^s \\
\pm_1^t-_1^s\pm_2^t+_2^s \\
\pm_1^t-_1^s\pm_2^t-_2^s \\
\pm_1^t+_1^s\mp_2^t+_2^s \\
\pm_1^t+_1^s\mp_2^t-_2^s \\
\pm_1^t-_1^s\mp_2^t+_2^s \\
\pm_1^t-_1^s\mp_2^t-_2^s
\end{array}
& 
\begin{array}{c}
(|s_1|\pm2)(|s_2|\pm2)/4 \\
3(|s_1|\pm2)|s_2|/4 \\
3|s_1|(|s_2|\pm2)/4 \\
9|s_1||s_2|/4 \\
(|s_1|\pm2)(|s_2|\mp2)/4 \\
3(|s_1|\pm2)|s_2|/4 \\
3|s_1|(|s_2|\mp2)/4 \\
9|s_1||s_2|/4 
\end{array}
\\ \hline
\end{array} \notag
\endgroup
\end{align}
Table 2: The number of zero-modes on each normal subspace for the $(T^2_1 \times T^2_2)/\mathbb{Z}_N$ twist and shift orbifolds. 
The first (second) column shows the types of orbifolds from $T^2_1$ ($T^2_2$), which also include $T^2_1$ ($T^2_2$) itself.
The third column shows the normal subspaces (eigenmodes) labeled by the $\mathbb{Z}_2$-twist eigenmodes $\pm^t$ and the $\mathbb{Z}_N$-shift eigenmodes $\pm^s$ for $(T^2_1 \times T^2_2)/\mathbb{Z}_N$.
The notation is same with~Table 1.
The last column shows the number of zero-modes on each normal subspace.


\subsection{$(T^2_1 \times T^2_2)/\mathbb{Z}_2$ permutation orbifold models}
\label{T2xT2imodels}

Next, we consider the $(T^2_1 \times T^2_2)/\mathbb{Z}_2$ permutation orbifolds. 
It is obtained by further identifying the complex coordinates of $T^2_1 \times T^2_2$, $(z_1,z_2)$ with the $\mathbb{Z}_2$ discrete interchanged points $I_2^n(z_1,z_2) \equiv e^{i\pi n} (z_{1+n},z_{2+n})$, $n \in \mathbb{Z}_2 = \{0,1\}$, $z_{\rm odd} \equiv z_1$, $z_{\rm even} \equiv z_2$. 
We can easily check $I_2^n \circ I_2^m = I_2^{n+m}$. 
In this identification, the wavefunctions on the magnetized $(T^2_1 \times T^2_2)/\mathbb{Z}_2$ permutation orbifolds, $\psi^{J,|M|}_{(T_1^2 \times T^2_2)/\mathbb{Z}_2^n}(z_1,z_2)$, are required to satisfy the following further boundary condition,
\begin{align}
\psi^{J,|M|}_{(T_1^2 \times T^2_2)/\mathbb{Z}_2^n}(I_2^1 (z_1,z_2))
= \psi^{J,|M|}_{(T_1^2 \times T^2_2)/\mathbb{Z}_2^n}(e^{i\pi} (z_{2}, z_{1}))
= e^{i\pi n} \psi^{J,|M|}_{(T_1^2 \times T^2_2)/\mathbb{Z}_2^n}(z_1,z_2), \quad n \in \mathbb{Z}_2, \label{interchangedT2xT2Z2base}
\end{align}
and, therefore, can be expressed by liner combinations of the wavefunctions on $T^2_1 \times T^2_2$ as 
\begin{align}
\psi^{J,|M|}_{(T_1^2 \times T^2_2)/\mathbb{Z}_2^n}(z_1,z_2) &= {\cal N}_2^p \sum_{\ell=0,1} (e^{i\pi n})^{-\ell} \psi^{j,|M|}_{0,T^2_1}(I_2^{\ell} (z_1),\tau) \psi^{k,|M|}_{0,T^2_2}(I_2^{\ell} (z_2),\tau),\ j \geq k,\ j,k \in \mathbb{Z}_{|M|},
\end{align}
where the normalization factor ${\cal N}_2^p = 1/2$ and $ 1/\sqrt{2}$ for $j=k$ and $j\neq k$, respectively. 
Note that there exists the magnetic flux condition $M_1 = M_2 \equiv M$ to identify two tori $T^2_1$ and $T^2_2$. 
Furthermore, the modulus parameters are also required to satisfy the condition $\tau_1 = \tau_2 \equiv \tau$, but it has been already assumed. 
There are the  $|M|(|M|+1)/2$ number of $\mathbb{Z}_2$-even modes ($n=0$) and $|M|(|M|-1)/2$ -odd modes ($n=1$). 
Under these liner combinations, the formula of modular forms  in Eq.~(\ref{modularformT2xT2}) becomes
\begin{align}
\psi^{(jk),|M|}_{(T_1^2 \times T^2_2)/\mathbb{Z}_2^n}(\gamma (z_1,z_2,\tau)) &= J_1({\gamma}, \tau) \sum_{m=0}^{|M|-1} \sum_{\ell = 0}^{m}
\rho_{(T^2_1 \times T^2_2)/\mathbb{Z}_2^n}({\gamma})_{(jk)(m\ell)} \psi^{(m\ell),|M|}_{(T_1^2 \times T^2_2)/\mathbb{Z}_2^n}(z_1,z_2,\tau).
\end{align}
The unitary representation $\rho_{(T^2_1 \times T^2_2)/\mathbb{Z}_2^n}$ is given by
\begin{align}
\rho_{(T^2_1 \times T^2_2)/\mathbb{Z}_2^n} (\gamma) _{(jk)(m\ell)} 
=  \left( \rho(\gamma)_{(jk)(m\ell)} + (-1)^n \rho(\gamma)_{(jk)(\ell m)} \right),\ j \geq k,\ m \geq \ell,
\end{align}
where it is multiplied by further factor $1/2$ for $m = \ell$. 
It satisfies the same algebraic relations with $\Gamma_{2|M|}'$. 
Thus, the representations on the $(T^2_1 \times T^2_2)/\mathbb{Z}_2$ permutation orbifold obey the same algebra with that on $T^2_1 \times T^2_2$, but their dimensions are smaller. 
For example, the $\mathbb{Z}_2$-permutation even modes on the $(T^2_1 \times T^2_2)/\mathbb{Z}_2$ permutation orbifold with the magnetic flux $M=2$ is given by
\begin{align}
\begin{pmatrix}
\psi^{(00),2}_{(T_1^2 \times T^2_2)/\mathbb{Z}_2^0}(z_1,z_2) \\
\psi^{(10),2}_{(T_1^2 \times T^2_2)/\mathbb{Z}_2^0}(z_1,z_2) \\
\psi^{(11),2}_{(T_1^2 \times T^2_2)/\mathbb{Z}_2^0}(z_1,z_2)
\end{pmatrix}
= 
\begin{pmatrix}
\psi^{0,2}_{0,T^2_1}(z_1) \psi^{0,2}_{0,T^2_2}(z_2) \\
\frac{1}{\sqrt{2}} \left(\psi^{1,2}_{0,T^2_1}(z_1) \psi^{0,2}_{0,T^2_2}(z_2) + \psi^{0,2}_{0,T^2_1}(z_1) \psi^{1,2}_{0,T^2_2}(z_2) \right)\\
\psi^{1,2}_{0,T^2_1}(z_1) \psi^{1,2}_{0,T^2_2}(z_2) 
\end{pmatrix}.
\end{align}
The unitary representations of the $S$ and $T$ are expressed as follows:
\begin{align}
\rho_{(T^2_1 \times T^2_2)/\mathbb{Z}_2^0} (S) = \frac{i}{2}
\begin{pmatrix}
1 & \sqrt{2} & 1\\
\sqrt{2} & 0 & -\sqrt{2} \\
1 & -\sqrt{2} & 1
\end{pmatrix},\quad
\rho_{(T^2_1 \times T^2_2)/\mathbb{Z}_2^0} (T) =
\begin{pmatrix}
1 & 0 & 0\\
0 & i & 0 \\
0 & 0 & -1
\end{pmatrix}.
\end{align}
These matrices are the same as those in Eq.~(\ref{eq:S4'-3}).
They generate the group $\Gamma'_4 \simeq S'_4$ which has the order 48.
The three zero-modes correspond to a triplet of $\Gamma'_4 \simeq S'_4$.
Thus, orbifolding by twist, shift, and permutation can decompose 
reducible representations into smaller one such as a reducible representation by their eigenvalues.


\subsection{$(T^2_1 \times T^2_2)/\mathbb{Z}_N$ twist, shift and permutation orbifold models}
\label{T2xT2tsimodels}

Now, we are ready to write down the algebraic relations for the unitary representations of the zero-mode wavefunctions on the $(T^2_1 \times T^2_2)/\mathbb{Z}_N$ twist, shift, and permutation orbifolds. 
As we saw in the previous subsection, the $\mathbb{Z}_2$ permutation does not affect the algebraic relations for the representations, although it is the permutation across the two tori $T^2_1$ and $T^2_2$. 
Thus, the algebraic relations in section 4.1 is valid for orbifolds including the $\mathbb{Z}_2$ permutation. 
Note that to identify two tori $T^2_1$ and $T^2_2$, the only pairs of the same $\mathbb{Z}_2$-twist ($\mathbb{Z}_N$-shift) eigenmodes are allowed on the $(T^2_1 \times T^2_2)/\mathbb{Z}_N$ twist, shift and permutation orbifolds.

As shown in the previous subsection, we can construct the wavefunctions on the $(T^2_1 \times T^2_2)/\mathbb{Z}_N$ twist, shift and permutation orbifolds as
\begin{align}
\Psi^{J,|M|}_{(T_1^2 \times T^2_2)/\mathbb{Z}_N^{(m;\ell_1,\ell_2;n)}}(z_1,z_2) 
= \frac{1}{\sqrt{2}} \sum_{p=0,1} (e^{i\pi n})^{-p} 
\psi^{j,|M|}_{T^2_1/\mathbb{Z}_{N}^{(m;\ell_1,\ell_2)}}(I_2^p (z_1),\tau) \psi^{k,|M|}_{T^2_2/\mathbb{Z}_{N}^{(m;\ell_1,\ell_2)}}(I_2^p (z_2),\tau),
\end{align}
where $\psi^{j,|M|}_{T^2/\mathbb{Z}_{N}^{(m;\ell_1,\ell_2)}}$ denotes the wavefunctions on the $T^2/\mathbb{Z}_N$ twist and shift orbifolds. 
The unitary representation $\rho_{(T^2_1 \times T^2_2)/\mathbb{Z}_N^{(m;\ell_1,\ell_2;n)}}$ is also given by
\begin{align}
\rho_{(T^2_1 \times T^2_2)/\mathbb{Z}_N^{(m;\ell_1,\ell_2;n)}} (\gamma)_{(jk)(pq)} 
&=  \left( \rho_{(T^2_1 \times T^2_2)/\mathbb{Z}_N^{(m;\ell_1,\ell_2)}} (\gamma)_{(jk)(pq)} 
+ (-1)^n \rho_{(T^2_1 \times T^2_2)/\mathbb{Z}_N^{(m;\ell_1,\ell_2)}} (\gamma)_{(jk)(qp)} \right), 
\end{align}
where
\begin{align}
\rho_{(T^2_1 \times T^2_2)/\mathbb{Z}_N^{(m;\ell_1,\ell_2)}} (\gamma)_{(jk)(pq)}
\equiv \rho_{T^2_1/\mathbb{Z}_N^{(m;\ell_1,\ell_2)}} (\widetilde{\gamma})_{jp}
\rho_{T^2_2/\mathbb{Z}_N^{(m;\ell_1,\ell_2)}} (\widetilde{\gamma})_{kq},
\end{align}
and $\rho_{T^2/\mathbb{Z}_N^{(m;\ell_1,\ell_2)}}$ denotes the representations on the $T^2/\mathbb{Z}_N$ twist and shift orbifolds. 
Then, as shown in Table 3, we can obtain the algebraic relations for the unitary representation on each orbifold. 
The dimension of each normal subspace (eigenmode) is shown in Table 4.

\begin{align}
\begingroup
\renewcommand{\arraystretch}{1.3}
\begin{array}{c|c|c} \hline
{\rm Orbifolds} & {\rm Magnetic\ flux} & {\rm Algebra\ for\ each\ mode} \\ \hline \hline
\mathbb{Z}_2\ {\rm permutation} & M_1 = M_2 = M & \Gamma_{2|M|}' \\ \hline
\mathbb{Z}_2\ {\rm twist}\ \&\ {\rm permutation} & M_1 = M_2 = M & \Gamma_{2|M|}'(\pm_1^t\pm_2^t) \\ \hline
\mathbb{Z}_N\ {\rm shift}\ \&\ \mathbb{Z}_2\ {\rm permutation}  & M_1 = M_2 = M = N^2s,\ s \in 2\mathbb{Z} & \left\{
\begin{array}{l}
\Gamma_{2|s|}'(+_1^s+_2^s)\\ 
\Gamma_{2|M|}'(-_1^s-_2^s) 
\end{array}
\right. \\ \hline
\mathbb{Z}_2\ {\rm twist\ \&\ shift} \ \&\ {\rm permutation} & M_1 = M_2 = M = N^2s,\ s \in 2\mathbb{Z} & \left\{
\begin{array}{l}
\Gamma_{2|s|}'(\pm_1^t\pm_2^t,+_1^s+_2^s)\\ 
\Gamma_{2|M|}'(\pm_1^t\pm_2^t,-_1^s-_2^s)
\end{array}
\right. \\ \hline
\end{array} \notag
\endgroup
\end{align}
Table 3: The algebraic relations for the unitary representation on the $(T^2_1 \times T^2_2)/\mathbb{Z}_N$ twist, shift and permutation orbifolds.
The first column shows the types of orbifolds. The second column shows the magnetic fluxes. The last column shows the algebraic relations for the unitary representation.
The notation is same with Table 1.

\begin{align}
\begingroup
\renewcommand{\arraystretch}{1.3}
\begin{array}{c|c|c} \hline
{\rm Orbifolds} & {\rm Normal\ subspaces} & {\rm \#\ of\ zero-modes} \\ \hline \hline
\mathbb{Z}_2\ {\rm permutation} & \pm^i & |M|(|M| \pm 1)/2 \\ \hline
\mathbb{Z}_2\ {\rm twist}\ \&\ {\rm permutation} & 
\begin{array}{c}
\pm_1^t\pm_2^t\pm^i \\
\pm_1^t\pm_2^t\mp^i
\end{array}
&
\begin{array}{c}
(|M|\pm2)(|M|\pm4)/8 \\
|M|(|M|\pm2)/8
\end{array}
\\ \hline
\mathbb{Z}_N\ {\rm shift}\ \&\ \mathbb{Z}_2\ {\rm permutation}  &
\begin{array}{c}
+_1^s+_2^s\pm^i \\
-_1^s-_2^s\pm^i
\end{array}
&
\begin{array}{c}
|s|(|s|\pm1)/2\\
|s|(N^2-1)(|s|(N^2-1)\pm1)/2
\end{array}
\\ \hline
\begin{array}{c}
\mathbb{Z}_2\ {\rm twist\ \&\ shift}\ \&\ {\rm permutation}
\end{array}
&
\begin{array}{c}
\pm_1^t+_1^s\pm_2^t+_2^s\pm^i \\
\pm_1^t+_1^s\pm_2^t+_2^s\mp^i \\
\pm_1^t-_1^s\pm_2^t-_2^s\pm^i \\
\pm_1^t-_1^s\pm_2^t-_2^s\mp^i
\end{array}
&
\begin{array}{c}
(|s|\pm2)(|s|\pm4)/8\\
|s|(|s|\pm2)/8\\
3|s|(3|s|\pm2)/8\\
3|s|(3|s|\mp2)/8
\end{array}
\\ \hline
\end{array} \notag
\endgroup
\end{align}
Table 4: The number of zero-modes on each normal subspace for the $(T^2_1 \times T^2_2)/\mathbb{Z}_N$ twist, shift and permutation orbifolds.
The first column shows the types of orbifolds.
The second column shows the normal subspaces labelled by the $\mathbb{Z}_2$-twist eigenmodes $\pm^t$, the $\mathbb{Z}_N$-shift eigenmodes $\pm^s$ and the $\mathbb{Z}_2$-permutation eigenmodes $\pm^i$ for $(T^2_1 \times T^2_2)/\mathbb{Z}_N$. 
The last column shows the number of zero-modes on each normal subspace.
The notation is same with Table 1.

\section{Conclusion}
\label{conclusion}

In this paper, we have discussed the modular symmetry on $(T^2_1 \times T^2_2)/\mathbb{Z}_N$ assuming the modulus parameters are identified. 
This identification allows us to regard the zero-mode wavefunctions on $(T^2_1 \times T^2_2)/\mathbb{Z}_N$ as the modular forms of weight 1. 
Moreover, the modular symmetry on $T^2_1 \times T^2_2$, $\Gamma \times \Gamma$, is broken to $\Gamma$.
Zero-modes are multiplets of the favor symmetry $\Gamma_{2{\rm lcm}(|M_1|,|M_2|)}'$ and $\Gamma_{2{\rm lcm}(|M_1|,|M_2|)}$ 
depending on orbifolding.
Only if we consider the pair of the $T^2/\mathbb{Z}_2$ twist orbifolds, the flavor symmetry of zero-modes are 
 $\Gamma_{2{\rm lcm}(|M_1|,|M_2|)}$.
In the case of other orbifolds obtained by the $\mathbb{Z}_2$ twist, the full $\mathbb{Z}_N$ shift and the $\mathbb{Z}_2$ permutation, the flavor symmetries are given by $\Gamma_{2{\rm lcm}(|M_1|,|M_2|)}'$.
Especially, we have shown the realization of the double cover of $S_4$, i.e., $S_4'$.
That would be interesting from the recent bottom-up approach of model building~\cite{Novichkov:2020eep,Liu:2020akv}.
Also the flavor symmetry $\Gamma_{2{\rm lcm}(|M_1|,|M_2|)}$ would be interesting.

Orbifolding decomposes zero-modes by eigenvalues of the $\mathbb{Z}_2$ twist, 
the $\mathbb{Z}_N$ shift and the $\mathbb{Z}_2$ permutation, 
and reduce the number of zero-modes, namely the generation number of quarks and leptons.
Three-generation models on twist orbifolds $T^2/\mathbb{Z}_2$ with magnetic fluxes 
have been classified in Refs.~\cite{Abe:2008sx,Abe:2015uma}.
Combinations of orbifolding by the $\mathbb{Z}_2$ twist, 
the $\mathbb{Z}_N$ shift and the $\mathbb{Z}_2$ permutation provide us 
with the further possibility to construct  three-generation models.
We would study such model building and its phenomenological aspects elsewhere.

\vspace{1.5 cm}
\noindent
{\large\bf Acknowledgement}\\

T. K. was supported in part by MEXT KAKENHI Grant Number JP19H04605. 
H. O. was supported in part by JSPS KAKENHI Grant Numbers JP19J00664 and JP20K14477.
H. U. was supported by Grant-in-Aid for JSPS Research Fellows No. 20J20388.





\end{document}